\newcommand\lam{\mbox{$\:\lambda\ $}}
\newcommand\lamlam{\mbox{$\:\lambda\lambda $ }}
\newcommand\ha{{H$\alpha$}}
\newcommand\hb{{H$\beta$}}

\newcommand\kms{\:\rm{km\,s^{-1}}}

\newcommand\FLUX{\:{\rm erg\:cm^{-2}\:s^{-1}}}

\newcommand\VEL{\:{\rm km\:s^{-1}}}

\newcommand\SiiL{[\ion{S}{2}] $\lambda\lambda 6716, 6731$}

\newcommand\sii{[\ion{S}{2}]}
\newcommand\nii{[\ion{N}{2}]}
\newcommand\oi{[\ion{O}{1}]}
\newcommand\oii{[\ion{O}{2}]}
\newcommand\oiii{[\ion{O}{3}]}
\newcommand\feii{[\ion{Fe}{2}]}
\newcommand\hii{\ion{H}{2}}
\newcommand\hi{\ion{H}{1}}

\newcommand\gal{NGC\,6946}

\newcommand{\EXPU}[3]{\mbox{\rm $#1 \times 10^{#2} \rm\:#3$}}  

\documentclass[modern]{aastex62}

\shorttitle{SNRs in NGC~6946}
\shortauthors{Long, Winkler, \& Blair}

\usepackage{natbib}
\usepackage{comment}
\usepackage{todonotes}
\usepackage{graphicx}
\begin{document}

\title{A New, Larger Sample of Supernova Remnants in NGC\,6946}

\correspondingauthor{Knox S. Long}
\email{long@stsci.edu}

\author[0000-0002-4134-864X]{Knox S. Long}
\affil{Space Telescope Science Institute,
3700 San Martin Drive,
Baltimore MD 21218, USA; long@stsci.edu}
\affil{Eureka Scientific, Inc.
2452 Delmer Street, Suite 100,
Oakland, CA 94602-3017}

\author[0000-0001-6311-277X]{P. Frank Winkler}
\affiliation{Department of Physics, Middlebury College, Middlebury, VT, 05753; 
winkler@middlebury.edu}

\author[0000-0003-2379-6518]{William P. Blair}
\affiliation{The Henry A. Rowland Department of Physics and Astronomy, 
Johns Hopkins University, 3400 N. Charles Street, Baltimore, MD, 21218; 
wblair@jhu.edu}



\begin{abstract}

The relatively nearby spiral galaxy NGC~6946 is one of the most actively star forming galaxies in the local Universe.  Ten supernovae (SNe) have been observed since 1917, and hence \gal\ surely contains a large number of supernova remnants (SNRs).  
Here we report a new optical search for these SNRs  using narrow-band images obtained with the WIYN telescope.  We
identify 147 emission nebulae 
as likely SNRs,  based on elevated \sii:\ha\ ratios compared to \hii\ regions.  We have obtained spectra of 102 of these  nebulae with Gemini North--GMOS; of these, 89 have \sii:\ha\ ratios greater than 0.4, the canonical optical criterion for identifying SNRs.  There is very little overlap between our sample and the SNR candidates identified  by Lacey et al. (2001) from radio data.  Also, very few of our SNR candidates are known X-ray sources, unlike the situation in some other galaxies such as M33 and M83.  The emission line ratios, e.g., \nii:\ha, of the candidates in \gal\ are typical of those observed in SNR samples from other galaxies with comparable metallicity.  
None of the  candidates observed in our low-resolution spectra show evidence of anomalous abundances or  significant velocity broadening.   A search for emission at the sites of all the  historical SNe in \gal\ resulted in detections for only two: SN\,1980K and SN\,2004et.  Spectra of both show very broad, asymmetric line profiles, consistent with the interaction between SN ejecta and the progenitor star's circumstellar material, as seen in late spectra from other core-collapse SNe of similar age.

\end{abstract}

\keywords{galaxies: individual (NGC~6946) -- galaxies: ISM  -- supernova remnants}



\section{Introduction} \label{sec:intro}

\gal\ is a nearby \cite[$6.72\pm 0.15$ Mpc,][]{tikhonov14}, nearly face-on \cite[$i = 32.6\degr$,][]{deblok08} galaxy with four flocculent spiral arms.  The galaxy is currently undergoing a major starburst, and as a result it has been the site of ten historical supernovae (SNe) since 1917, the most of any known galaxy. According to \citet{jarrett13} the total star formation rate is $3.2\, M_\sun\,{\rm yr}^{-1}$, a high rate that is thought to be bar-driven.  A total of 121 bubbles, probably created by stellar winds and multiple SNe at the star-formation sites, have been identified in \hi\ gas that extends well outside the bright portions of the optical galaxy \citep{boomsma08}.  
Given these properties, one expects that a large number of supernova remnants (SNRs) should be present in \gal, since SNRs tend to remain visible for thousands of years.   

Optically, SNRs are usually identified on the basis of high \sii:\ha\ ratios compared to \hii\ regions.  In bright \hii\ regions, most sulfur is found in the form of S$^{++}$ \citep[or above, e.g.,][]{pagel78, levenson95}, and as a result the \sii:\ha\ ratios are typically 0.1 or smaller.  In SNRs, where emission is driven by impulsive heating from shock waves, S$^+$ is found in an extended recombination and cooling zone behind the shock, and the \sii:\ha\ ratios are typically $\gtrsim0.4$.\footnote{There are SNRs, including Tycho's SNR and SN1006, that have optical spectra dominated by Balmer line emission and with  little if any \sii, but they are rare \cite[see, e.g.][]{heng10}. All are thought to be the products of SN Ia explosions. All are young SNRs expanding into a tenuous ISM with shock velocities that are so high that a recombination zone has not had time to develop.   Such SNRs would be missed in the type of survey we describe here.  As such objects age, they should become detectable.}  The diagnostic can become less deterministic as one pushes to lower surface brightness, as recently discussed by \cite{long18} for the case of M33.

The first optical search for SNRs in \gal\ was made by \cite{matonick97}, hereafter MF97, who used interference filter imagery to identify 27 emission nebulae with \sii:\ha\ ratios $\geq$ 0.45 as SNRs.  One of these sources, MF-16, was later associated with the ultraluminous black hole X-ray binary \gal\ X-1 \citep{roberts03}. Though very rare, such ultra-luminous X-ray sources (ULXs) have hard X-ray spectra that produce line ratios in the surrounding circumstellar ISM that resemble those expected from SNRs. To our knowledge, no other optical searches for SNRs in NGC~6946 exist, nor have spectra of the remaining MF97 objects ever been reported.  


Here we discuss a new, more sensitive optical search for SNRs in \gal\ in which we identify a total of 147 SNR candidates using interference filter imagery.  We also discuss spectroscopic observations of 102 of these candidates, which we use to verify the ratios obtained from the imaging and to characterize other characteristics of our new optical sample.  The paper is organized as follows:  Section 2 describes both our imaging and spectroscopy observations,  presents our significantly expanded catalog of optical SNR candidates, and the results of our spectroscopy.  Section 3 discusses these results in the context of both \gal\ and other spiral galaxies, and Section 4 presents our detection and late-time spectra of two of the historical SNe in \gal. Finally, Section 5 provides a brief summary of our results.

\section{Observations and Data Reduction \label{sec:observations}}
\subsection{Imaging and Catalog of SNR Candidates}

We carried out narrow-band imaging observations of \gal\ from the 3.5m WIYN telescope and MiniMosaic imager on Kitt Peak on the nights of 2011 June 26-28 (UT).\footnote{The WIYN Observatory is a joint facility of the University of Wisconsin-Madison, Indiana University, the National Optical Astronomy Observatory and the University of Missouri.}  The so-called ``Minimo'' was mounted at the f/6.3 Nasmyth port and consisted of a pair of $2048\times4096$ SITe chips, with a field  9\farcm 6 square at a scale of 0\farcs 14 pixel$^{-1}$.  We used interference filters that pass lines of \ha, \sii \lamlam6716,6731, and \oiii \lam 5007, plus red and green narrow-band continuum filters so we could subtract the stars and produce pure emission-line images. Frames in each filter were dithered to enable automatic removal of cosmic rays and bad pixels.  Further observational details are given in Table 1.  

It is noteworthy that the \ha\ filter was quite narrow in bandwidth, 27\ \AA\ FWHM. Its transmission is 69\% at the rest wavelength of \ha, but only 11\% at 6548\ \AA\ and 16\% at 6583\ \AA; hence, the \nii\ lines are greatly attenuated relative to \ha.\footnote{The recessional velocity of \gal\ is only $40 \kms$, so the lines are redshifted by $\lesssim 1$ \AA. Also, lab measurements of the filters confirm only small shifts in centroids due to the f/6.3 beam.}  This facilitates identification of SNRs based on their image-derived \sii:\ha\ ratios.  Seeing throughout this run was about $1\arcsec \approx 32$ pc at the distance of \gal.  While sufficient for identifying SNRs in all but the most crowded regions, this resolution was insufficient for obtaining properties such as diameter or morphology.

\begin{figure}
\plotone{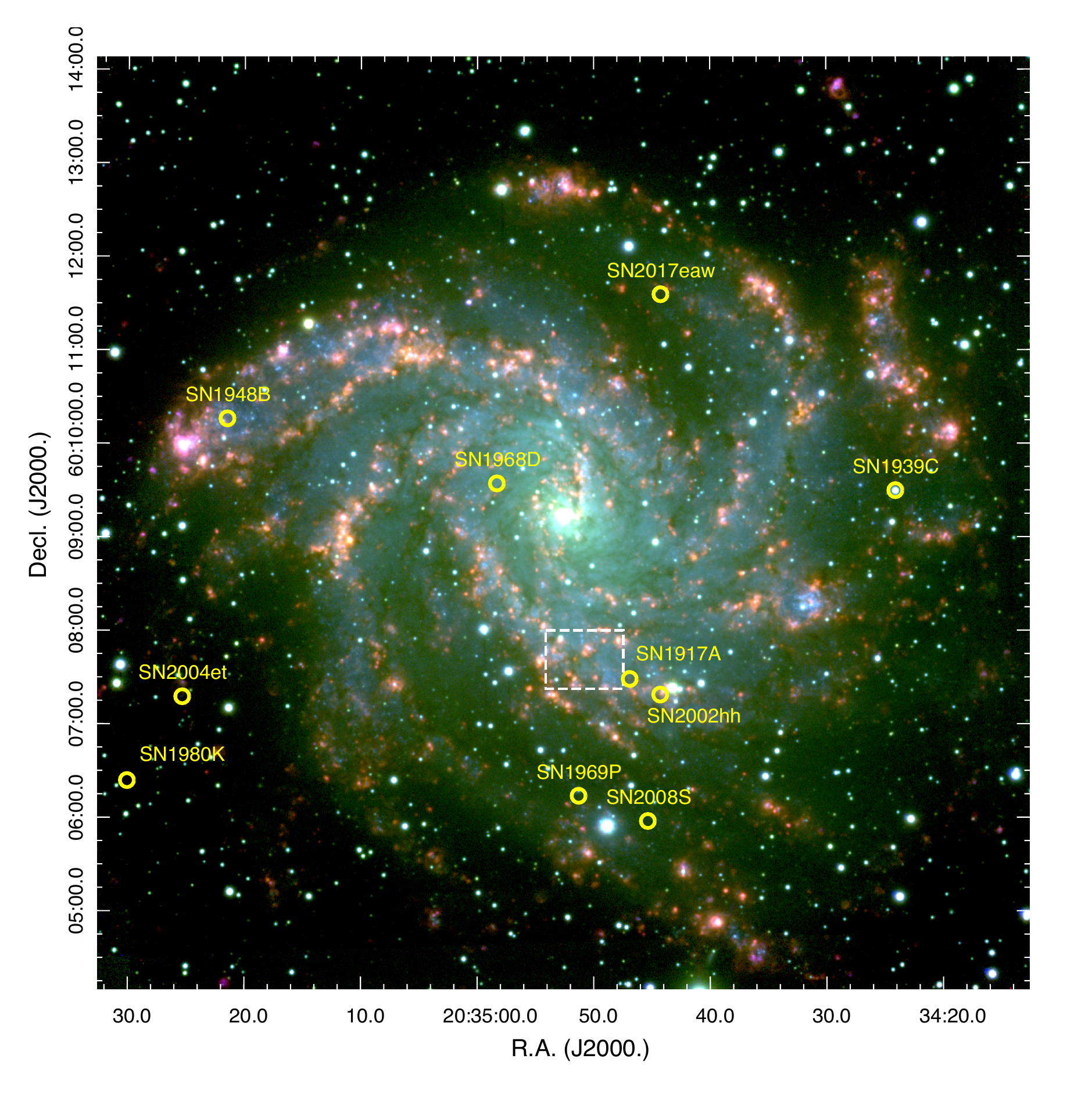}
\caption{An emission-line image of \gal, where R = \ha, G = \sii, and B = \oiii, taken from the 3.8m WIYN telescope at Kitt Peak.  This figure uses the emission line images prior to continuum subtraction, so the underlying galaxy light is also visible. Yellow circles indicate the positions of the ten historical SNe since 1917 (including SN\,2017eaw, which occurred subsequent to our observations). The dashed rectangle indicates the small region shown in detail in Fig.~\ref{fig_diagnostic}.  The field is 10\arcmin\ square. 
\label{fig_rgb_overview}}
\end{figure}

We used standard IRAF\footnote{IRAF is distributed by the National Optical Astronomy Observatory, which is operated by the Association of Universities for Research in Astronomy, Inc., under cooperative agreement with the National Science Foundation.} techniques for processing the images, including overscan correction, bias subtraction, and flat-fielding using dome flats.  Procedures in the IRAF {\texttt mscred} package were then used to combine the data from the individual chips into a mosaic image for each frame, assigned a WCS for each using stars from the USNOB1 catalog \citep{monet03}.  We then stacked  all the images by filter onto an arbitrary standard coordinate system with a scale of 0\farcs 20 pixel$^{-1}$ and scaled and subtracted the continuum images from the emission-line ones (red from \ha\ and \sii; green from \oiii) to give pure emission-line images with most of the stars and galaxy background light removed.  Finally, we used observations of seven different spectrophotometric standard stars---all selected for their weak \ha\ absorption lines---from the catalog of \citet{massey88} to flux calibrate the emission-line images.  Figure \ref{fig_rgb_overview} shows a color version of the final images (R = \ha, G = \sii, B = \oiii), with the positions of historical SNe in \gal\ also indicated.

\begin{figure}
\plotone{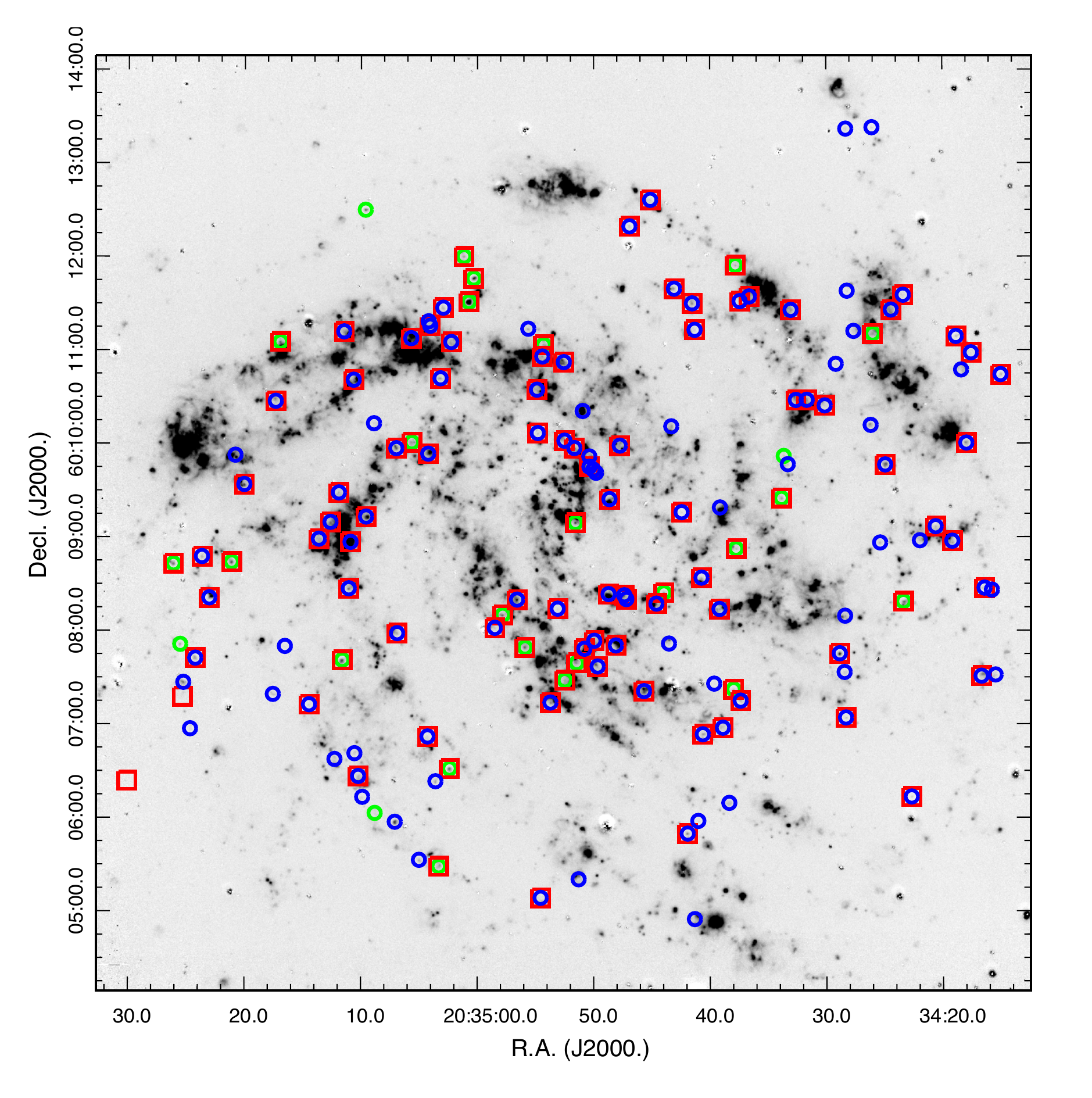}
\caption{Image of \gal\ in \ha, after continuum subtraction to remove most of the stars.  Green circles indicate the positions of SNRs and candidates from MF97; blue circles indicate the positions of our new \sii-selected candidates.  The red squares denote the subset of objects for which we obtained GMOS spectra (Table~4).  The field is identical to that shown in Fig.~\ref{fig_rgb_overview}. \label{fig_index_overview}}
\end{figure}

To select SNR candidates, we used the DS9 display program to show the continuum-subtracted WIYN images in all three emission lines as well as a \sii:\ha\ ratio image and a continuum image (to discriminate stars or stellar subtraction residuals from point-like nebulae).   We then visually inspected these to select SNR candidates based on a high \sii:\ha\ ratio.  The initial inspection was carried out by Middlebury undergraduate Marc DeLaney; subsequently two of us (WPB and PFW) compiled independent candidate lists; we then conferred to agree on a consensus list of 147 candidates, including the 27 from MF97. The positions of all 147 candidates are shown in Figure ~\ref{fig_index_overview}. The vast majority of the SNR candidates appear within the spiral arms or clustered on the outskirts of large complexes of \hii\ emission and star formation.

\begin{figure}
\plotone{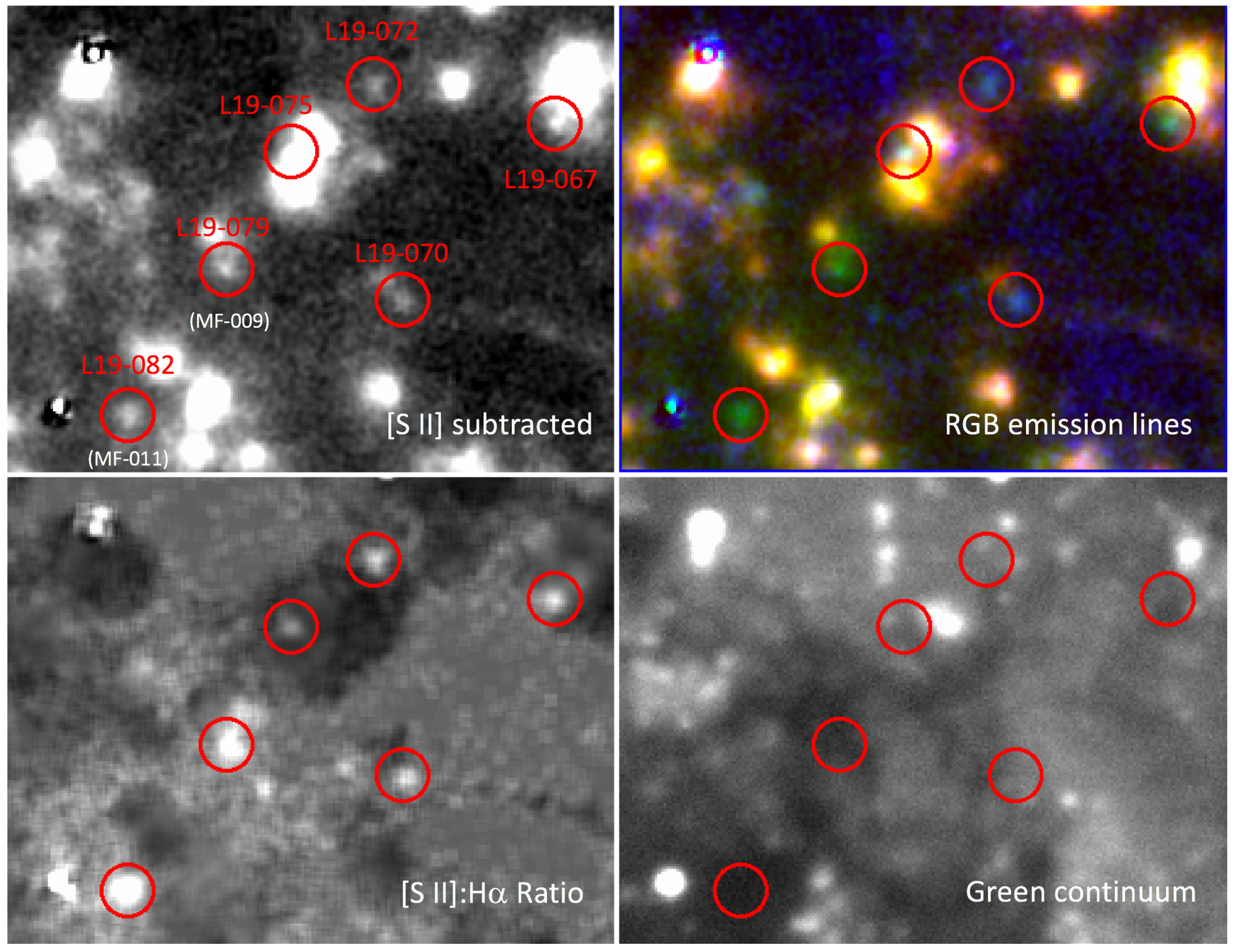}
\caption{This figure demonstrates the diagnostic process used to find SNR candidates in \gal.  The region shown is 35\arcsec\ in the N-S dimension and is centered $\sim$1.5\arcmin\ south of the nucleus (see Fig. 1).  At upper left is a continuum-subtracted \sii\ image for reference.  At upper right, we show a color image of subtracted emission line images, where red is H$\alpha$, green is \sii, and blue is \oiii.  Bottom left shows the \sii:H$\alpha$ ratio image of the region, where elevated values of the  ratio are white and low values are black.  The lower right image shows the green continuum image, which is useful for identifying the presence of stellar subtraction residuals.  The red circles are 4\arcsec\ in diameter and show two MF97 objects and four newly identified SNR candidates from our survey (identifications shown in upper left panel).  Note the appearance in the ratio image, where most of the emission nebulae show low ratios, but the objects in the red circles stand out in the ratio.  In the upper right panel, SNR candidates appear as greenish compared to photoionized nebulae, due to relatively stronger \sii\ and/or \oiii\ emission. \label{fig_diagnostic}}
\end{figure}

Figure \ref{fig_diagnostic} shows an example for a small region $\sim$1.5\arcmin\ south of the nucleus (see Fig. ~\ref{fig_rgb_overview}) containing two of the MF97 objects and several new SNR candidates. In this example we have combined the three continuum-subtracted emission line images into a single color panel that shows how the stronger \sii\ and/or \oiii\ emission from the SNR candidates makes them stand out.  The \sii:\ha\ ratio image was a key diagnostic for drawing our eyes to potential objects of interest. Then an assessment of the integrity of the candidate as an emission object was made by inspection of the actual images.  This was done to separate candidates from stellar residuals or false regions of higher ratio in the ratio map that were due to moise.

Having performed similar SNR searches in other galaxies such as M33 and M83, it is worth noting some differences for \gal.  M33 is of course much closer, and the nebulae of interest are almost always resolved.  M83 is much more distant (4.6 Mpc) than M33, but our search there was aided by the exceptional 0.5\arcsec\ seeing conditions we obtained at the Magellan telescope \citep{blair12}.  Many objects were resolved, but others extended down to the limits of what even HST could resolve ($\sim$1-2 pc) \citep{blair14}.  Our WIYN survey of \gal\ did not have exceptional seeing, and the distance is some 50\% larger than for M83.  Hence, relatively few of the nebulae of interest are resolved, and with variations in the complex galaxy background, it is much more difficult to perform a systematic search.  The use of the \sii:\ha\ ratio image was particularly helpful for \gal, which is the most distant galaxy for which we have performed this kind of ground-based SNR survey. Nonetheless, while we have expanded the SNR candidate list very substantially, clearly completeness has not been achieved.  Our list of 147 SNR candidates falls far short of M83 for example (with 300+), even though the SN rate is larger in \gal.  Higher resolution data (better seeing and/or HST imaging) would no doubt help substantially, but the greater distance for \gal\ is still a limiting factor.

In addition, we initially selected 51 emission nebulae with relatively high \oiii:\ha\ ratios (using an \oiii:\ha\ ratio image in the display).  Most of these nebulae were expected to be \hii\ regions but we hoped that one or more might be a young, ejecta-dominated SNR, simlar to Cas A in our Galaxy. With ten SNe in the last century and a high incidence of massive stars, one might expect a number of young, ejecta-dominated SNRs to be present.  Though none of the O-selected candidates had ratios as extreme as seen for Cas A or 1E0102-72.3 in the Small Magellanic Cloud, we nevertheless selected some of them for follow-up spectroscopy.  None of these nebulae for which we obtained spectra have any indication of ejecta in their spectra or the broad lines would expect from a very young SNR.  All are \hii\ regions with somewhat enhanced ionization state, and so these form the bulk of the \hii\ sample we use for comparison with the SNR candidates sample below. Finally, we inspected the positions of the nine\footnote{As of the 2011 observations there were nine SNe; a tenth SN was recorded in 2017.} historical SNe in \gal\ for evidence of nebular emission; we only detected line emission at the positions of SN1980K and SN2004et, both of which we targeted for follow-up spectroscopy. (See section 3.4 below.)

The SNR candidates are listed in Right Ascension order in Table \ref{table_candidates}.  For each candidate, we provide (1) a source name, (2,3) the position  (J2000), (4) the deprojected galactocentric distance (GCD), (5) the \ha\ flux as derived from the emission line images, (6) the \sii:\ha\ ratio measured from the images, (7) the spectrum we used to confirm the imaging ratios (see below), (8) whether or not the object has a spectroscopic confirmation that the  \sii:\ha\ ratio is $\ge$ 0.4 (see below) and (9) other names for the source.

\subsection{Spectroscopy; Emission-Line Fluxes}

We used the Gemini Multi-Object Spectrograph (GMOS) on the 8.2m Gemini-North telescope to obtain all the spectra reported here, during queue-scheduled programs in semesters 2014B (program GN-2014B-Q-83) and 2015B (program GN-2015B-Q-91).  
For the 2014B program, we designed six custom masks, each with 20-30 slitlets targeting SNR candidates whose positions we determined from our 2011 WIYN images, together with  short $R$-band pre-images of several \gal\ fields taken with GMOS earlier in 2014 as part of the spectroscopy preparation program.    
We used two additional masks (which we refer to as masks 7 and 8 for simplicity)  for the 2015B program.  Slitlets in one or more of our eight masks were placed on 102 distinct SNR candidates, including ones with a range of sizes, GCDs, and ISM environments (locations in arms and in 
inter-arm regions).  In addition to the SNR candidates, we also placed a number of slitlets on \hii\ regions for comparison purposes in both 2014 and 2015.

We used the 600 line\,mm$^{-1}$ grating designated G5307 and a GG455 cut-off filter to block second-order spectra.   The detector in both years was a mosaic of three e2v deep-depletion CCD chips, binned by 2 in the spatial direction (for a scale of 0\farcs 146 pixel$^{-1}$) and by 4 in the dispersion direction.  
The dispersion was 1.84 \AA\,  pixel$^{-1}$ (binned), resulting in coverage of the spectral range from at least \hb\ through \sii\lamlam 6716,\,6731 for virtually all the objects.\footnote{The detailed wavelength coverage for individual objects naturally varied with slitlet position on the mask in the dispersion direction.}  Our  masks had  slitlet  widths from 1\farcs 25 to 1\farcs 75, with wider slits used for the larger objects, and lengths of 6\arcsec\ or longer to permit local background and sky subtraction.   

With each mask, we took spectra at three or four slightly different grating tilts, to cover wavelength gaps between chips and to gain somewhat more total spectral range.\footnote{An exception was mask 6, done late in the 2014B semester, for which our full set of planned observations were never completed.}  At each wavelength setting, we obtained two or more identical exposures to minimize the effects of cosmic rays.
For calibration, we programmed quartz flats and CuAr arc frames immediately before or after the science exposures with each mask and grating setting.  A journal of all the science observations from both 2014 and 2015 appears in Table \ref{obs_log}.  The SNR candidates for which we obtained spectra are indicated by the small red boxes in Fig.~\ref{fig_index_overview}.


The data were processed using standard procedures  from the {\tt gemini} package in IRAF  for bias subtraction, flat-fielding, wavelength calibration, and combination of spectra with different grating tilts to provide the final results.  
Flux calibration was based on baseline GMOS observations of a few spectrophotometric standard stars, carried out in the same semester as part of standard GMOS operations.  

During the processing, the 2-D spectra from different slitlets were separated to give individual 2-D spectra from each slitlet.   We examined each of these individually and selected the object region, as well as one or more sky background regions,  stripped out 1-D spectra of each, and then subtracted the sky spectrum from the corresponding object to obtain the final background-subtracted object spectra.  
Many of the objects are located in regions with bright surrounding galactic background (both continuum and emission lines) from NGC\,6946, so the selection of a representative local background was done on a best-effort basis.    In addition to the targeted SNR candidates and \hii\ regions, we extracted spectra from other \hii\ regions which appeared by chance in the slits when this was possible.   This allowed us to increase our sample of \hii\ regions from 24 to 65.
We then performed  fits to obtain emission line fluxes  from the spectra, assuming Gaussian profiles, for the following lines and line complexes:  \hb\ alone,  the \oiii\ doublet,  the \oi\ doublet,  the \ha-\nii\ region, and   the \sii\ doublet. For the fits, we assumed that the background varied linearly with wavelength around each line, and that the FWHM of all lines in each complex was the same.

\begin{figure}
\plotone{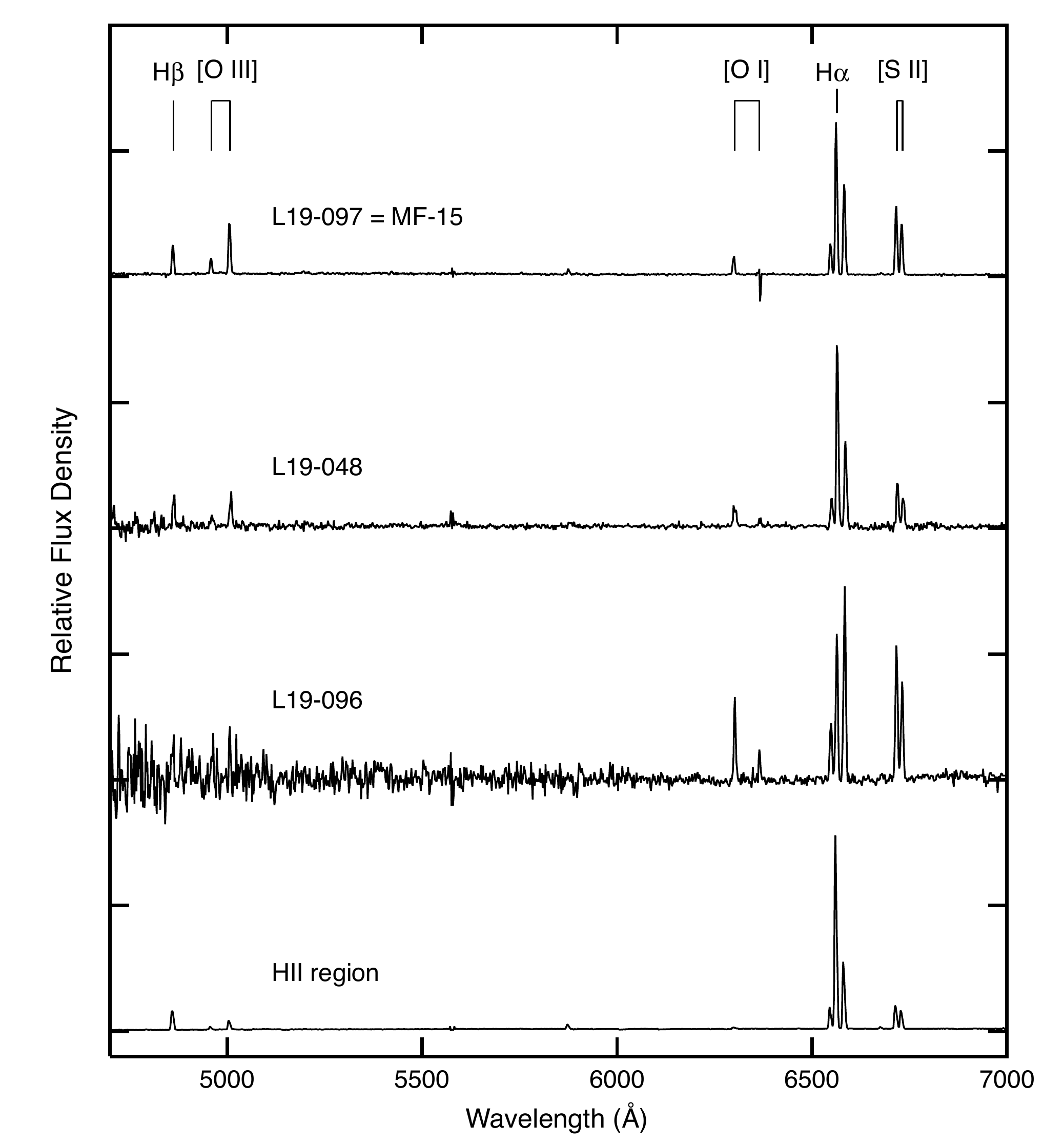}
\caption{Examples of the spectra obtained for three representative SNR candidates and one \hii\ region.  The three SNR spectra have been selected to illustrate the quality of the spectra for a bright, medium bright and fairly faint candidate. The traces have been scaled arbitrarily, and offset for clarity. \label{fig_example_spectra}}
\end{figure}

Representative examples of the spectra that were obtained are shown in Fig.\ \ref{fig_example_spectra}.   The three SNRs were selected primarily to show how the quality of the spectra changes as a function of brightness.  The spectroscopically obtained \sii:\ha\ ratio for L19-048 was 0.45, just above the value for spectroscopic confirmation, while those for the previously known bright object L19-097=MF-15 and the faint candidate L19-096 are higher.  All three SNR candidates show evidence of emission from [O I], which is another indicator that the emission we see arises from shock-heated gas. 

Table \ref{table_snr_spec}  lists the information we obtained for the SNR candidates for which we obtained spectra.  Specifically we list (1) the source name, (2) the extracted  \ha\ flux, (3-9) ratios of various emission lines to \ha\ [taken to be 300], (10) the total \sii:\ha\ ratio and (11) the measured FWHM of the lines in the \ha-\nii\ complex.  For doublets, where the line ratio is constrained by atomic physics, that is \oiii, \oi, and \nii, we have listed only the stronger line.  We visually inspected all of the spectra and the fits to them; values which we judge to be more uncertain are indicated with tildes in the Table.  No allowance has been made for additional errors associated with difficulties in background subtraction.  A number of the objects were observed with more than one mask, sometimes with different slit orientations.  In these cases, we used the spectrum which we judged to be the most accurate and report it for reference in the `Spectrum' column in Table \ref{table_candidates}, where (for example) 05.18 should be interpreted as `mask 5 slitlet number 18.'

\section{Analysis and Discussion}

\begin{figure}

\plottwo{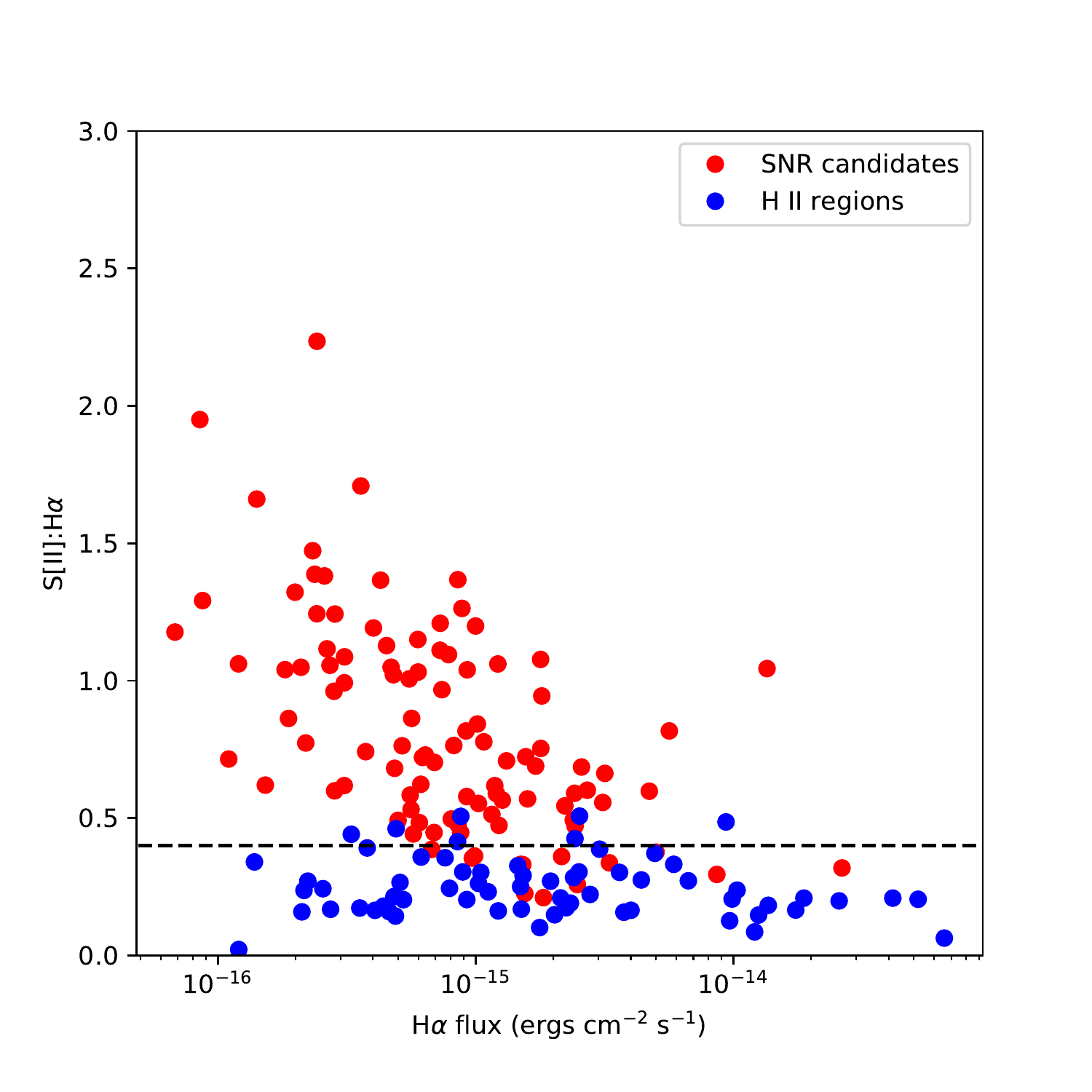}{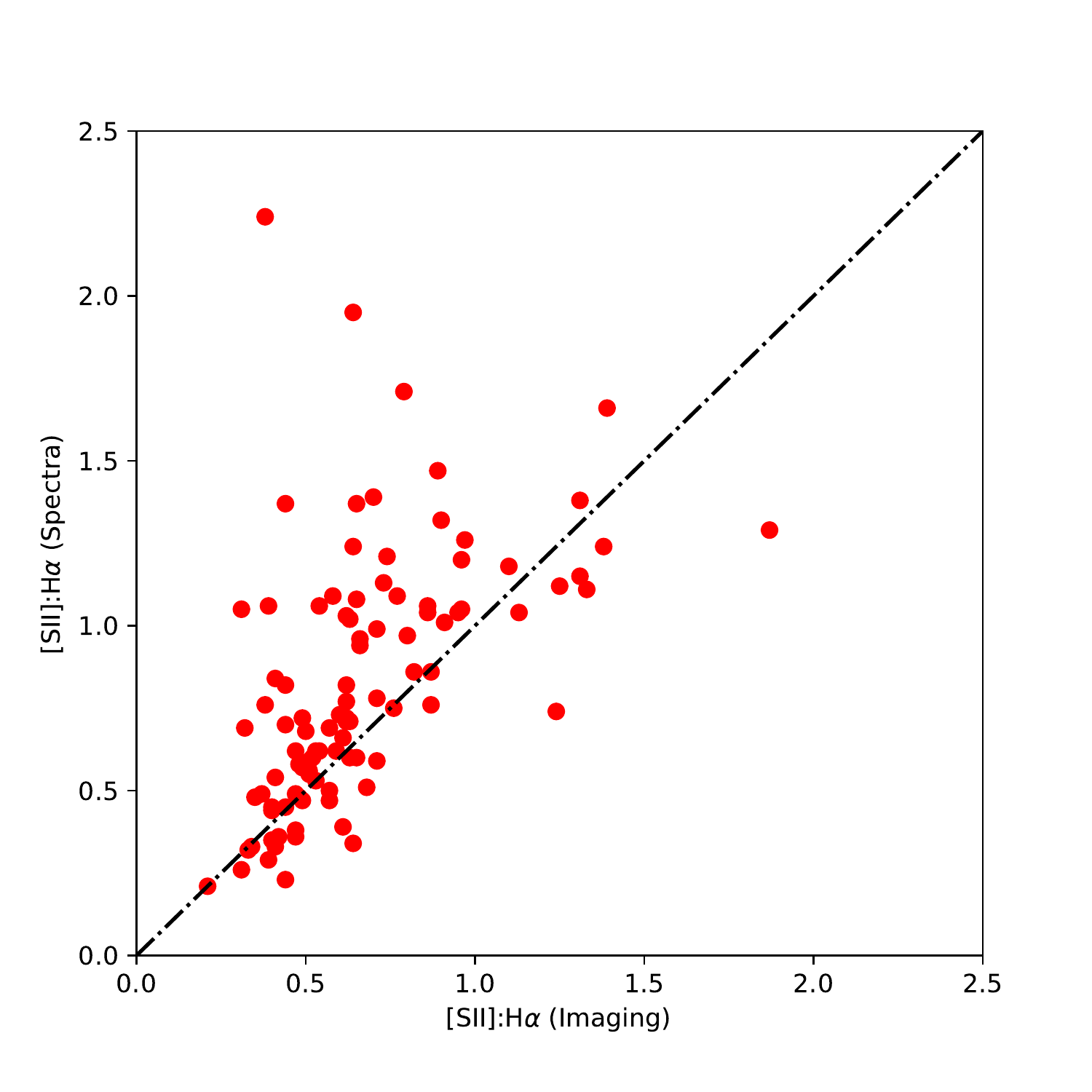}
\caption{Left: The \sii:\ha\ ratio obtained spectroscopically for SNR candidates (red) and \hii\ regions (blue) as a function of \ha\ flux in the spectrum.  Objects with ratios greater than 0.4 (the dashed line) are spectroscopically confirmed SNRs.  Objects near the dividing line are less certain because observational errors in the ratio and/or \hii\ contamination can impact the derived ratio. Right: The \sii:\ha\ ratio derived from spectra compared to that derived from narrow band imaging.  The tendency for spectroscopic ratios to be somewhat higher is consistent with mild impacts from \nii\ emission getting through the \ha\ imaging filter.  \label{fig_s2_ha}}

\end{figure}

Of the 147 candidate SNRs from the WIYN interference images, we obtained spectra of 102. The spectra were needed to improve our confidence that the \sii:\ha\ ratios of the emission nebulae were indeed high and to begin to characterize the SNRs using accurate line ratios.  The spectroscopic ratios are shown as a function of \ha\ flux in the left panel of  Fig.\ \ref{fig_s2_ha}. \hii\ regions, including both those selected for their \oiii\ emission and those that appeared serendipitously along spectral slits, are shown in blue, while nominal \sii:\ha\ candidates from imaging are shown in red.   Conventionally, emission nebulae are identified as SNRs optically if the \sii:\ha\ ratio exceeds 0.4; 89 of the 102 SNRs with spectra satisfy this criterion, and so we regard these as confirmed SNRs (and they have been listed as such in Table \ref{table_candidates}). 

Clearly, given uncertainties in derived line ratios for faint emission line objects, a dividing line of 0.4 is somewhat arbitrary, and objects just above (or just below) this ratio should be judged with more context.  For example, slightly under-subtracting contaminating \ha\ emission in the spectra could readily explain why some candidates ended up below the threshold in the spectroscopic analysis. We have inspected the objects listed with imaging ratios above the threshold and spectral ratios below, and indeed, many of them are located in regions of \hii\ contamination. Likewise, a slight over-subtraction of \ha\ could enhance the \sii:\ha\ ratio derived, potentially pushing some objects above the threshold.  This is likely the reason why some of the faint \hii\ regions observed spectroscopically actually lie above the 0.4 threshold.  The observed tendency to see higher \sii:\ha\ ratios closer to the nucleus is likely to be at least partly an abundance effect; similar trends are seen in M33 \citep{long18} and in M83 \citep{winkler17}. 

There are 45 objects without spectra, so which of these are actually SNRs is uncertain. Given the generally good agreement between imaging and spectral ratios, those objects with imaging ratios well in excess of 0.4 are likely to be good candidates. As shown in the right panel of Fig.\ \ref{fig_s2_ha}, if anything the spectroscopically-determined ratios tend to be higher than the ratios determined from narrow band imaging, and so most of the objects without spectra are likely to be SNRs.  The somewhat higher spectroscopic ratios are to be expected, since even with our relatively narrow \ha\ filter, some emission from \nii\ was  also passed.   

\subsection{Comparison to MF97}

MF97 identified 27 SNR candidates in \gal. All of these objects have \sii:\ha\ ratios in our WIYN images that exceed 0.4.  MF97 obtained spectra of six of their candidates.  We have obtained spectra of 23 of the MF97 objects, including new spectra of four objects for which MF97 had spectra -- MF-03, MF-21, MF-22, and MF-26.  All of these indeed have measured spectroscopic ratios that exceed 0.4.  We note that MF97 used a relatively conservative value of \sii:\ha\ $\ge$ 0.45 to establish their catalog (to avoid the issue of errors in the determined ratio affecting objects just above or below the normal 0.4 criterion).  Hence, it is perhaps not surprising that the previous objects are strongly confirmed here.  MF97 report ``typical 1.5\arcsec\ seeing'' for their work. (MF97 looked at a number of galaxies and they do not give a specific value of seeing for \gal.) Hence, it is also clear that MF97 were only able to find a combination of objects that were relatively bright (so not smeared out by seeing) and/or were relatively isolated from contaminating emission.  They estimated that at least four times as many SNRs were likely present in \gal, and our current survey has far surpassed that estimate.

Even though we have identified many more candidates than MF97, it is clear that with its better seeing and higher sensitivity, our survey  is still limited and likely to be significantly incomplete. For example, Fig.~\ref{fig_diagnostic} shows two objects, L19-067 and L19-075,  both in close proximity to \hii\ contamination,   that we were able to identify as candidates while MF97 could not.  However, it is not hard to imagine any number of additional objects in the many even more confused emission regions that our survey would have  missed.

The object MF-16, listed here as L19-098, deserves separate mention as it is far and away the brightest object in our catalog.  Originally thought to be a possible example of an exceedingly bright (and possibly multiple) SNR \citep{{blair94},{blair01}}, similar to the extraordinary SNR in NGC~4449 \citep{{blair83},{dannym08}}, X-ray variability was subsequently established that clearly indicates the presence of an accreting black hole binary within the nebular complex \citep{{roberts03}, {fridriksson08}, {rao10}}.  
Most recent analyses \citep{{kaaret10},{ berghea12}}  model the system based on the ULX binary only, but clearly the system involves some combination of shock-heated and X-ray photoionized emission.  \citet{dunne00} show resolved line profiles on the bright emission lines indicating kinematic motions of order 250 -- 400 $\rm km ~ s^{-1}$, and while \citet{roberts03} show the bulk of the X-ray emission is likely due to the ULX, they estimate the SNR component could be as bright as $\rm \sim ~ 2.5 ~ \times ~ 10^{38} ~ ergs ~ s^{-1}$ in X-ray, which is quite substantial for a SNR.  
Although jets are often invoked for accreting ULX binaries, the HST images of the nebula \citep{blair01} are not obviously consistent with this idea; the morphology shows a multiple loop structure, and the likely optical counterpart of the ULX is not centered in the smallest, brightest loop.  Hence, the idea that the complex involves something more complicated than a single SN that created the ULX binary may still be relevant to consider in understanding the overall characteristics of this intriguing object.

\subsection{Global Spectroscopic Properties of the SNR Candidates}

\begin{figure}
\plotone{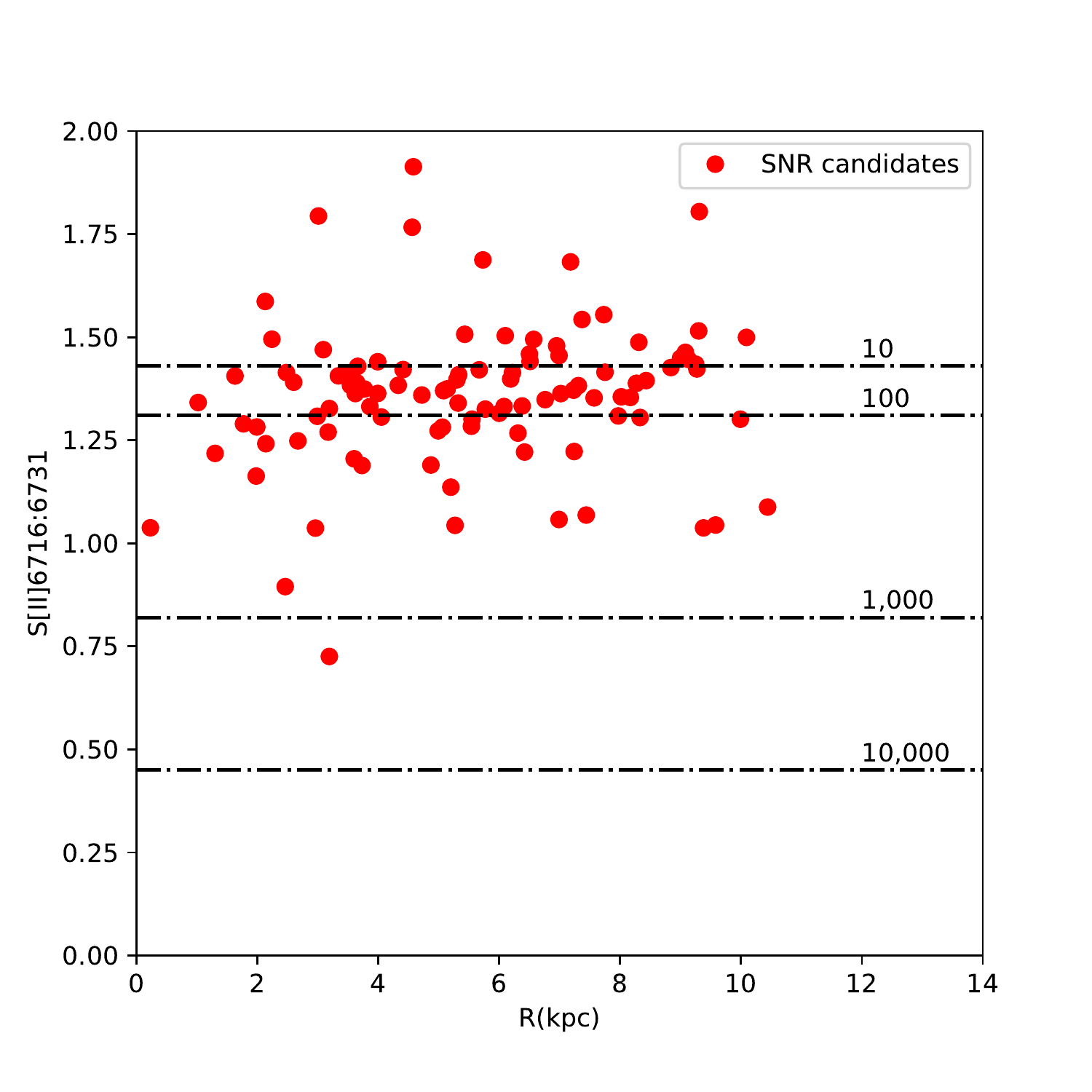}
\caption{The [S II]6716:[S II]6731 line ratio for SNR candidates as a function of galactocentric distance.  This ratio is nominally a density diagnostic, and most candidates are near the low density limit of 1.4, indicating generally low ISM densities. Derived values above 1.4 are non-physical, and are indicative of errors in the derived ratio.  Only a handful of objects appear to show densities significantly above the low-density limit.  There is no obvious trend with galactocentric distance.  \label{fig_s2_ratio}}
\end{figure}

As shown in Fig.\ \ref{fig_s2_ratio}, the density-sensitive \sii\ ratio  $\lambda$6716:$\lambda$6731 clusters around the low-density limit of 1.4 for the SNR candidates, and  the fact that  about as many objects have non-physical ratios above 1.4 as below suggests that a) the ratios for some faint objects have significant errors (not unexpected), and b) likely  almost all the objects are close to this limit. This is in contrast to the situation in M83 \citep{winkler17} and to a lesser extent in M33 (L18) where a significant number of the SNRs show evidence of high densities, especially for smaller diameter objects. In the absence of good SNR diameters here, we cannot search for trends with diameter, but upcoming {\em Hubble Space Telescope} observations should provide accurate diameter information for many of these objects.  These {\em HST} images could also reveal the presence of very small SNR candidates, perhaps with high densities, which eluded detection in our ground-based images.

\begin{figure}
\plotone{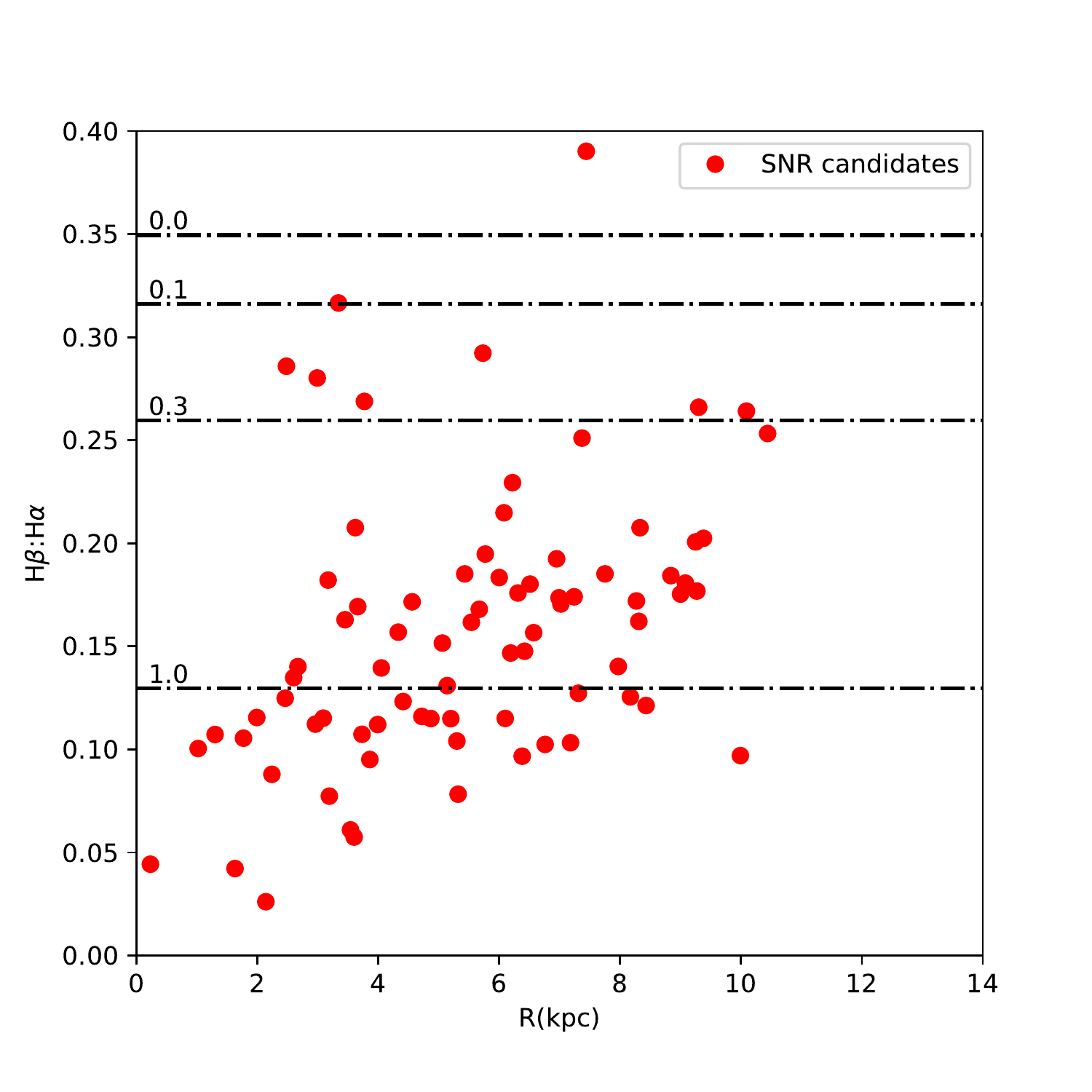}
\caption{Observed \hb:\ha\ line ratios for SNR candidates as a function of galactocentric distance. The dashed lines indicate the expected values of the line ratio for E(B-V) of 0.0, 0.1, 0.3 and 1.0.  Objects near the galactic center tend to be more highly reddened than those far from the center. \label{fig_reddening}}
\end{figure}

Fig.\ \ref{fig_reddening} shows the observed \hb:\ha\ ratios of the SNR candidates with spectra as a function of GCD. Nearly all of the SNRs show significant reddening, as one would expect since, at 12\degr\ from the Galactic plane, foreground reddening from within our Galaxy along the line of sight is expected to be $E(B-V) = 0.29$ \citep{schlafly11}.  There is clearly substantial internal and differential reddening  within \gal\ as well, as there is a very evident trend for objects near the center of \gal\ to be more reddened than those more distant from the nucleus.

\begin{figure}
\plottwo{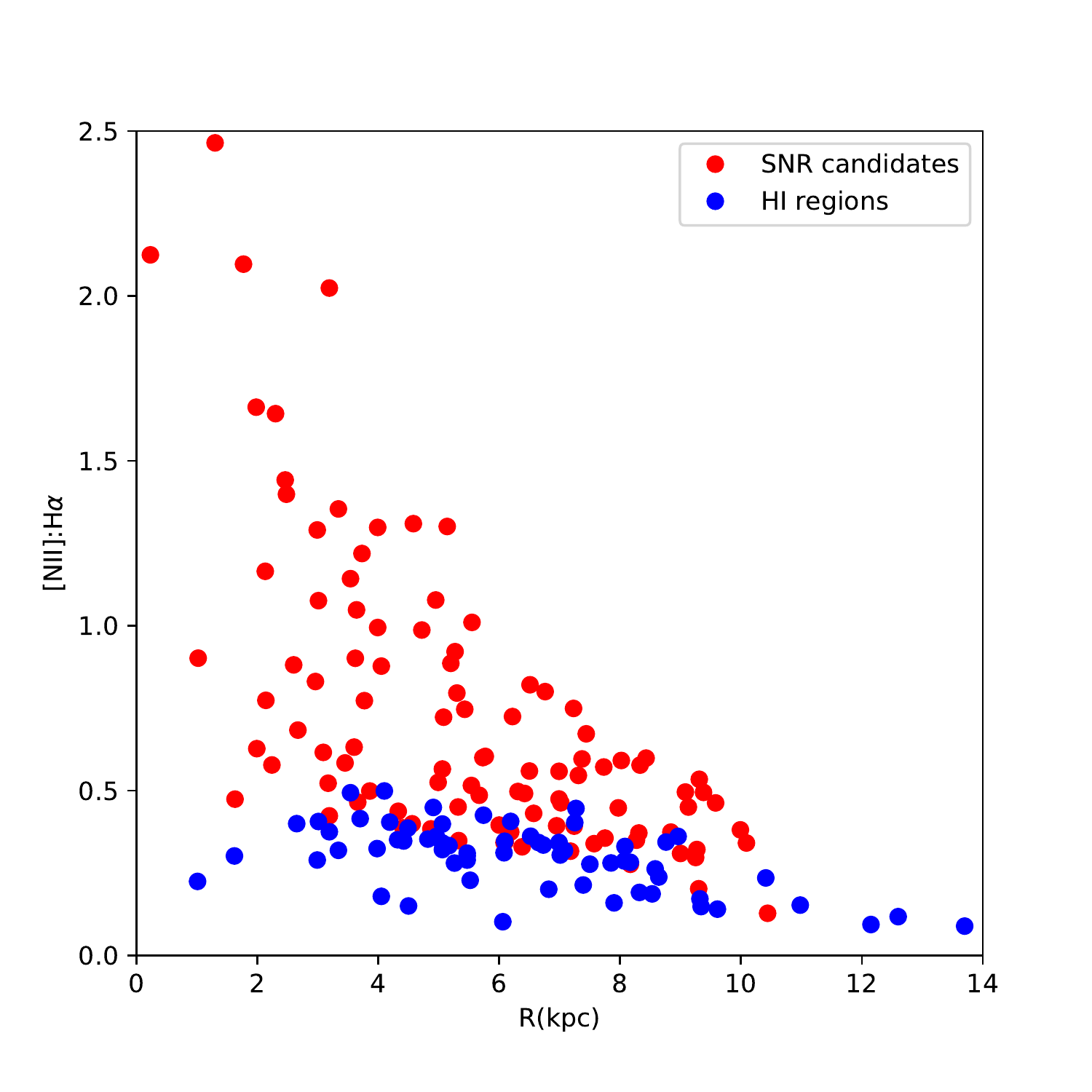}{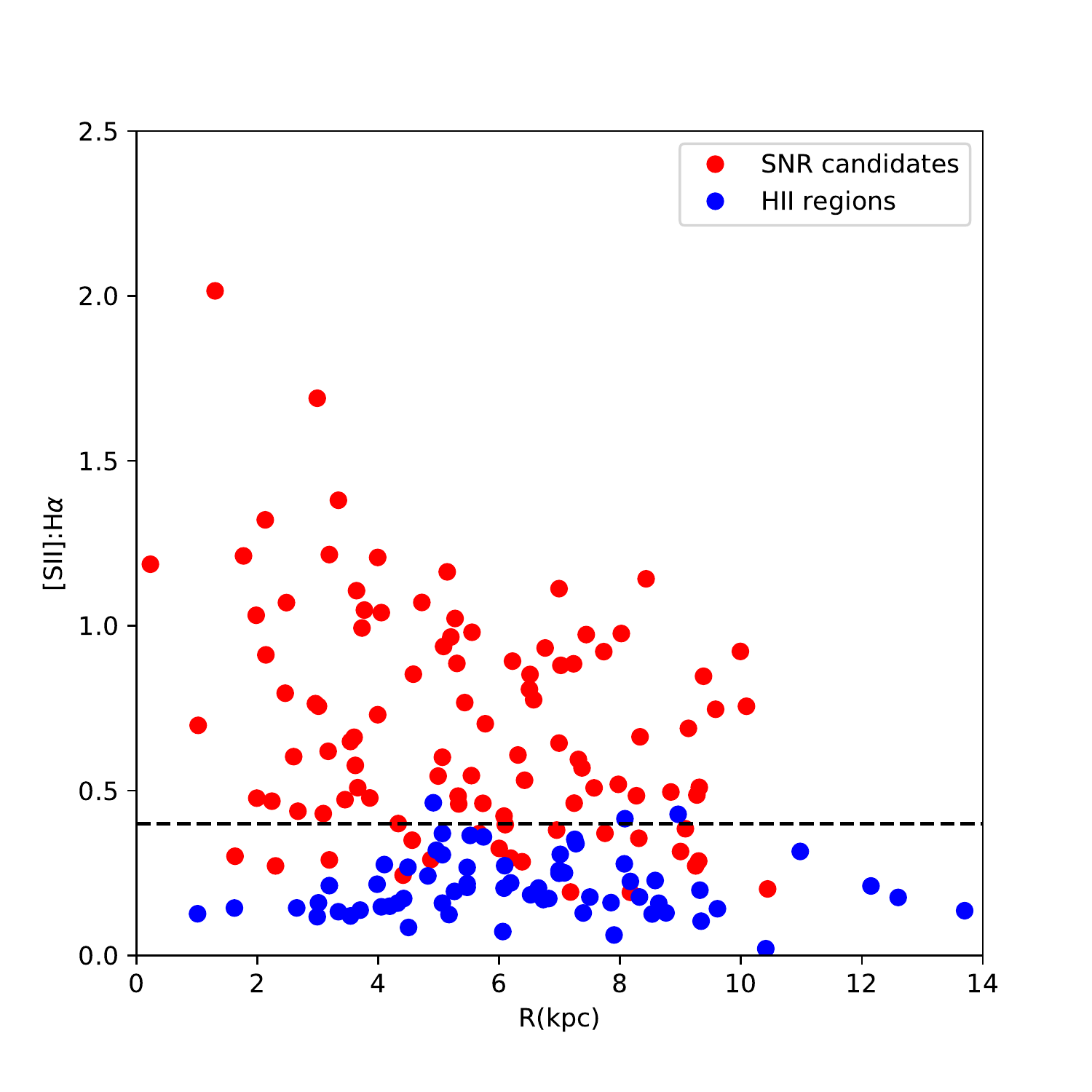}
\caption{Line ratios of [N II]:\ha\ and [S II]:\ha\ as a function of GCD.  Both SNR candidates and \hii\ regions are shown.     \label{fig_metal_ratios}}

\end{figure}

\begin{figure}
\plotone{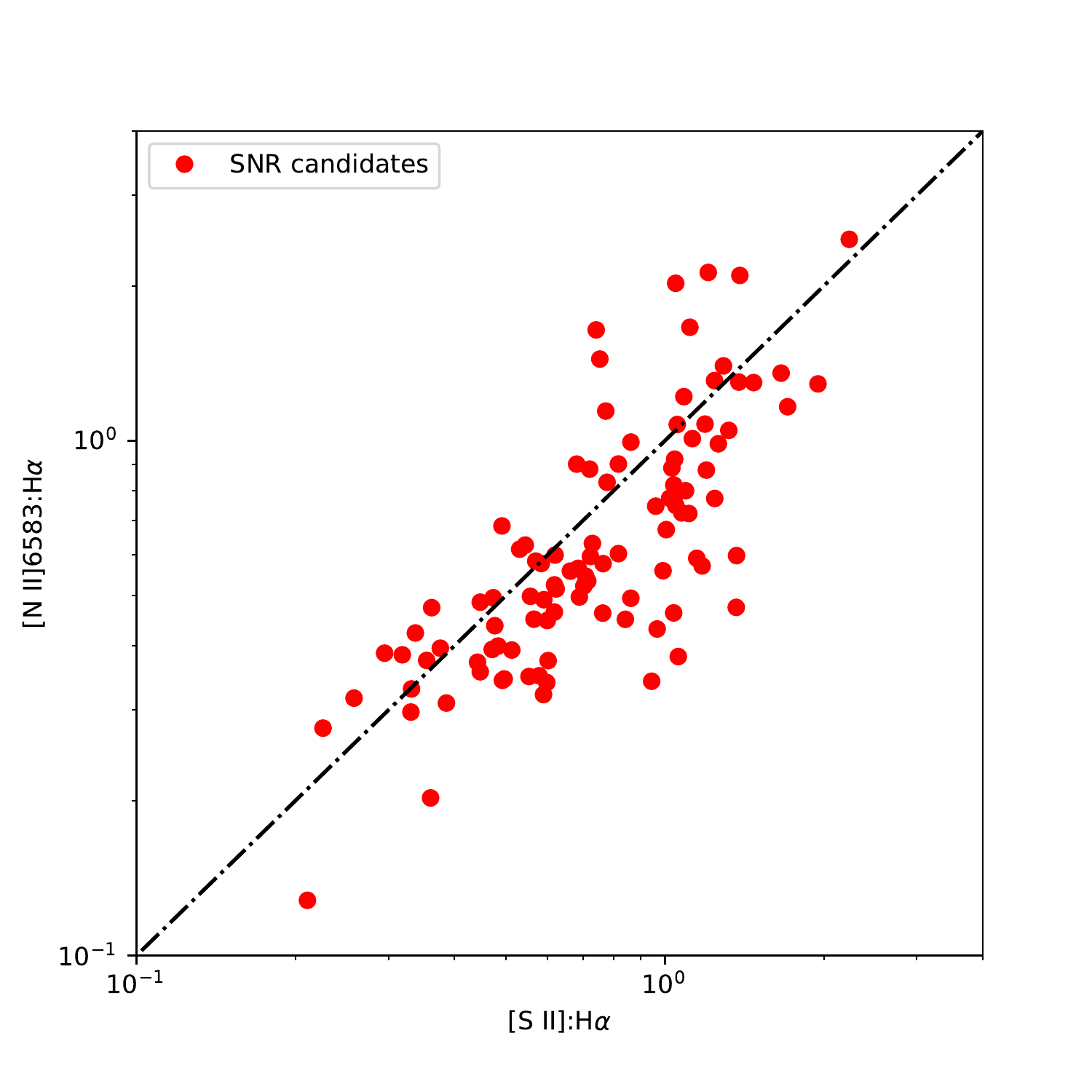}
\caption{A comparison of the line rations of [N II]6583:\ha\ to that of [S II]:\ha\ for the SNR candidates. \label{fig_s2_n2}}
\end{figure}

Our SNR line ratios show a  general decrease with increasing GCD as seen in Fig. \ref{fig_metal_ratios}, although the dispersion at any particular GCD distance is large.  The trend could well be indicative of abundance gradients in nitrogen and sulfur, with the dispersion being due to varying shock conditions and/or or local abundance variations.  However, the trend stands in contrast to SNRs in M33, where both the \nii:\ha\ and \sii:\ha\ ratios have a large range and do not decrease systematically with increasing GCD (although the line ratios in the \hii\ region sample appear well behaved---see L18 Fig. 8).   There is a good correlation between the \nii:\ha\ and \sii:\ha\ ratios of the various objects, as shown in  Fig. \ref{fig_s2_n2}, as also seen in other galaxy samples (cf. Winkler et al. 2017, Fig. 9).

\begin{figure}
\plottwo{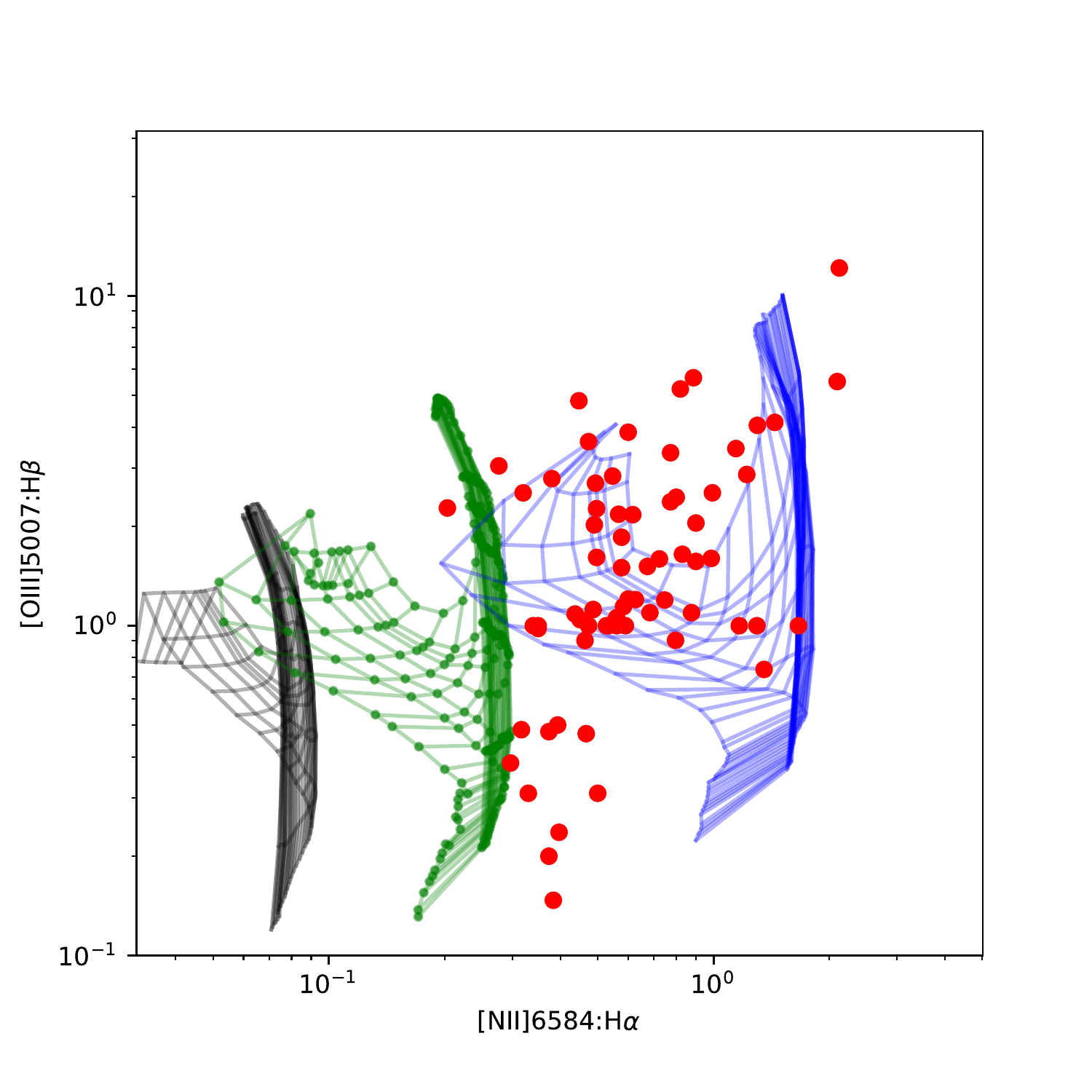}{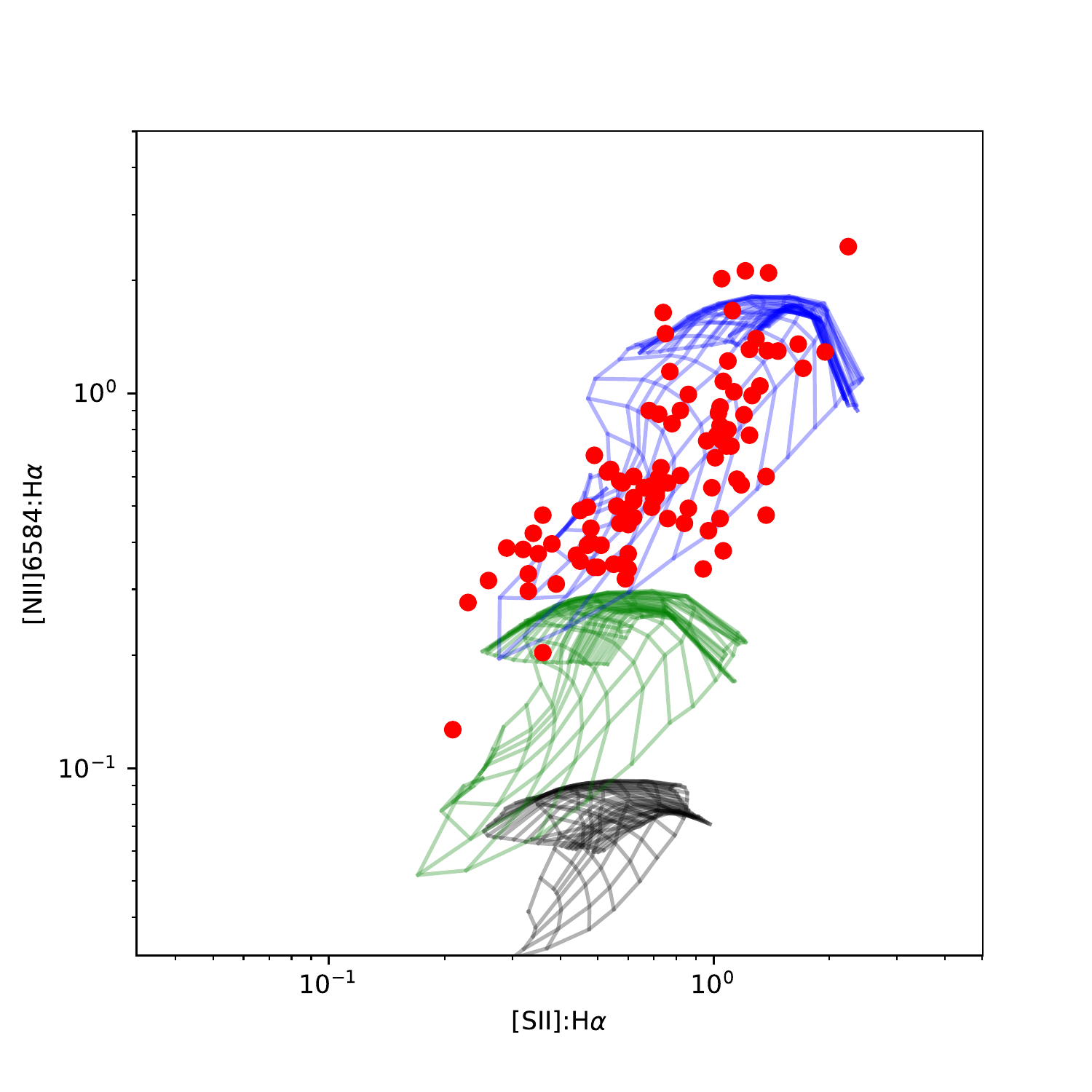}
\caption{Left: Model [O III] 5007:\hb\ ratio as function of the \nii\ 6583:\ha\ line ratio for SNRs and SNR candidates with spectra. As discussed in the text, the black, green and blue meshes correspond to shock models from Allen et al. (2008) with a range of shock velocities and pre-shock magnetic fields, and with metallicities corresponding to the SMC (black), LMC (green), and Milky Way (blue).  Right: Model \nii\ 6583:\ha\ line ratios as a function of the \sii:\ha\ ratios.  \gal\ objects appear consistent with solar abundances and a wide range of other physical properties.  \label{fig_model}}
\end{figure}

Line ratios in SNR spectra  are expected to vary both as a function of shock conditions and metallicity.  To see where the SNRs in \gal\ lie, we have compared the line ratios calculated from models by \cite{allen08} using the MAPPINGS III code for a range of shock velocities (100 - 1000 $\VEL$) and pre-shock magnetic fields (10$^{-4}$ - 10 $\mu$G).  The results are shown in Fig.~\ref{fig_model}.  The results both for the ratio of \oiii\,$\lambda$5007:\hb\ and for \sii:\ha\ fall squarely into the region of the solar metallicity models. This is consistent with expectations for previous abundance studies of \gal\ such as \citet{cedres12}, \citet{gusev13}, and references therein, depending of course on the adopted method of determining \hii\ region abundances from strong-line data only. Both of these papers also show a very modest abundance gradient in the \hii\ regions of \gal, with considerable scatter about the mean, although very few \hii\ regions are sampled in the inner portion of the galaxy (cf. {Cedr{\'e}s} et al. Fig. 18).

\subsection{SNRs in Other Wavelength Bands}

Searches for SNRs in \gal\ have been carried out in several other wavelength ranges.  In particular,  \cite{lacey01} identified 35 radio sources as SNR candidates on the criterion that these sources had non-thermal spectral indices and were positionally coincident with \ha\ emission.  These objects are located mostly in the spiral arms of \gal\ where there is active star formation and where one might expect SNe to explode.  There are seven emission nebulae in our list of SNR candidates that lie within 2\arcsec\ of radio SNR candidates; this is five more than had been identified previously, but a small fraction of the total number of radio objects.  \cite{lacey01} argued that the reason that few radio SNRs are detected optically is that the bright \ha\ emission from H II regions makes optical searches for SNRs less sensitive in the spiral arms than in the rest of the galaxy.  The fact that more of the optical SNRs are not detected at radio wavelengths is most likely a question of sensitivity.  All of the SNR candidates identified by \cite{lacey01} have radio luminosities of at least one-tenth of Cas A, much greater than the bulk of SNRs known in the Galaxy.  By contrast, in M33, where \cite{white19} have recently conducted a very deep radio survey with the Jansky Very Large Array, more than three-quarters of the optically identified SNRs have been detected at radio wavelengths.

An alternative diagnostic (to  the \sii:\ha\ ratio) for identifying SNRs in at least some external galaxies is emission in the \feii\  1.64 $\mu$m line.  Since Fe$^+$ is so easily ionized further, \hii\ regions are expected to have little if any \feii\, while \feii\ should, like \sii, be elevated in the cooling tail behind SNR shocks.  Hence, detection of an emission nebula with \feii\ is a strong indication of shock heating. In M83, where {\em HST} WFC3 IR imaging in \feii\ is available \citep{blair14}, about 40\% of the optical SNRs in the observed region were detected in \feii, and a handful of compact \feii\ nebulae in particularly dusty regions are strong SNR candidates whose emission is too highly absorbed to be detected optically. This raises the possibility that \feii\ might be valuable  not only to help confirm optical SNR candidates, but also to  help to obtain a more complete sample in heavily reddened regions.

In \gal, \citet{bruursema14}  carried out ground-based interference filter imagery in the light of \feii\ 1.64 $\mu$m. Ground-based \feii\ imaging is difficult due to sky contamination, as noted by  \citet{bruursema14}, but they were able to identify 48 candidate objects that they felt were above the noise.  Interestingly, only three of these objects align with SNR candidates in our sample: L19-076, L19-095=MF-14, and the exceedingly bright ULX MF-16 = L19-098 (discussed above). We are thus left wondering whether the other \feii\ objects are possible SNR candidates or whether the data quality issues are responsible for the large difference between \gal\ and M83. We can say, however, that the bulk of the \citet{bruursema14} candidates are not seen in projection onto the dustiest regions. M33, for which \citet{morel02} clearly detected a solid handful of optical SNRs in ground-based \feii\ observations, appears to be intermediate between the extremes of \gal\ and M83.  High spatial resolution {\em HST} WFC3 IR observations would make a large impact in clarifying the situation for a galaxy as distant and as highly absorbed as \gal.

SNRs are also X-ray sources, and therefore we have looked to see how many X-ray sources in \gal\ could be found in our candidate lists.  The most detailed X-ray study of \gal\ to date was carried out using {\em Chandra} by \cite{fridriksson08}, who constructed a catalog of 90 point sources, of which 25 appeared to be time variable (and hence likely X-ray binaries or background AGN).  Of the 90 point sources, there are eight which are positionally coincident with objects in our sample.  X-ray hardness ratios reported by \cite{fridriksson08} show that most of these have relatively soft X-ray spectra, consistent what is expected for thermal emission from a SNR.  The main exception is F08-08, coincident with L19-029, which has a hardness ratio that is more typical of X-ray binaries and background galaxies.  Not surprisingly, one of the X-ray sources coincident with L19 objects is  the ULX L19-098 = MF-16.  These two are also the only X-ray sources coincident with SNR candidates that also show evidence of (long-term) variability, according to \cite{fridriksson08}. 

Higher percentages of the optically identified sources have been X-ray-detected in M33 \citep[112/200,][]{long18} and M83 \citep[67/225,][]{long14}.  Of these, M83 is the more relevant for comparison.  M83, a nearly face-on grand-design spiral, has a star formation rate of 3-4 $M_\sun\,{\rm yr}^{-1}$ \citep{boisser05}, similar to \gal, but lies a distance of 4.61 Mpc \citep{sahu06} compared to 6.7 Mpc for \gal.  M83 was observed for 725 ks with {\em Chandra}, compared to a total of about 175 ks for \gal.    In addition, \gal\ is relatively close to the Galactic plane and as a result foreground absorption reduces the X-ray sensitivity, particular below 1 keV.  Indeed, the hydrogen column density along the line of sight to M83 is  \EXPU{4}{20}{cm^{-2}}, whereas  for \gal\ it is \EXPU{1.8}{21}{cm^{-2}} \citep{kaberla05}. For a thermal plasma with an effective temperature of 0.6 keV, the combination of greater distance and higher absorption implies that a typical SNR in \gal\ would have only about 1/3 the count rate of one in M83.  Consequently, it is not surprising that we have detected fewer SNRs in X-rays in \gal\ than in M83.

\section{Historical Supernovae in \gal}
\begin{figure}
\plotone{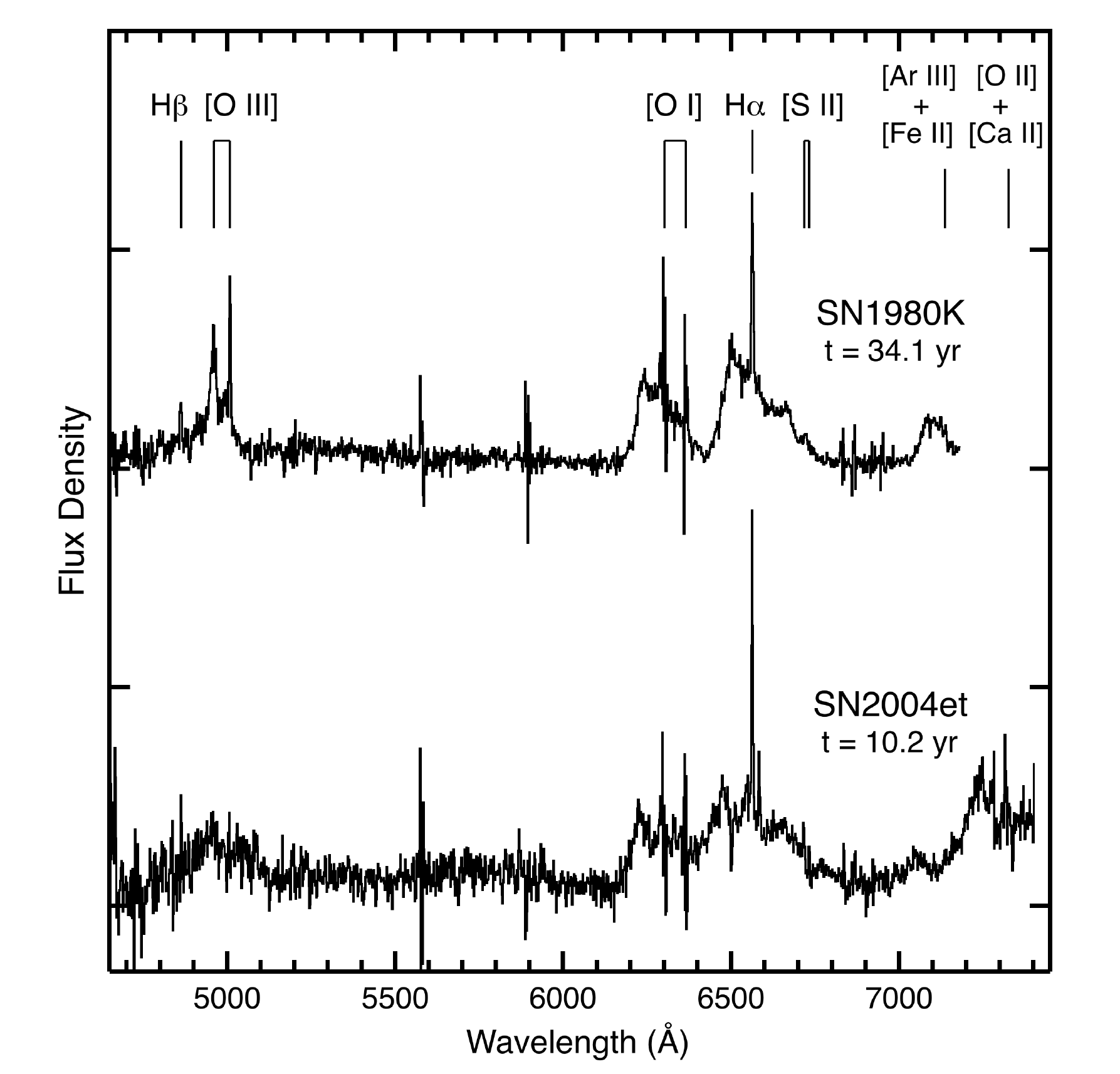}
\caption{GMOS spectra of the two historical SNe we have recovered in \gal.  Both show the broad emission lines characteristic of young core-collapse SNRs where rapidly expanding ejecta interact with a circumstellar shell.  (The spectra have been displaced vertically for clarity.) \label{historical_SNe}}
\end{figure}

In our 2011 WIYN emission-line images, we also searched for emission at the positions of all of the nine historical SNe that had occurred in \gal\ at the time of our observations.  We detected emission from only two of these: SN\,1980K and SN\,2004et.  Both would probably have been among our \sii-selected SNR candidates, and SN\,1980K would also have attracted notice because of its relatively high \oiii:\ha\ ratio as well; however, we noted these in an explicit search of the positions of all the historical SNe in \gal.  Thus, we have not included them in Tables \ref{table_candidates} or \ref{table_snr_spec}.
We  obtained GMOS spectra of both, as shown in Fig.~\ref{historical_SNe}.  Unlike the spectra from the other SNRs in our sample, the lines from both of these are highly velocity-broadened, the signature of fast shock waves in these young SN-SNR transition objects and fast-moving ejecta.

The Type IIL SN\,1980K has been frequently observed over the years since its explosion, and its transition from late-time SN to a developing SNR has been monitored both photometrically and spectroscopically \citep[e.g.,][]{uomoto86, fesen90, fesen94, fesen99b, milisavljevic12}.  Our 2014 GMOS spectrum, taken 3 December 2014---about  34 yr past maximum light, is qualitatively similar to the 30 yr spectrum shown by \citet{milisavljevic12}, with broad, asymmetric lines---stronger on the blue side than the red---from \ha, \oi, \oiii, and a feature near 7100 \AA\ that is probably [\ion{Fe}{2}]\,$\lambda$\,7155, possibly blended with [\ion{Ar}{3}]\,$\lambda$7136, all appearing above a faint, blue continuum \citep[see][for a discussion of the 7100 \AA\ feature]{fesen94}.  It appears that broad \SiiL\ with a similar asymmetric profile may also be present, blended with the red side of the  \ha\ line.  We estimate that the broad \ha\ flux is $8.4 \pm 1.0 \times 10^{-16} \FLUX$, slightly lower than that of $10 \pm 2 \times 10^{-16} \FLUX$ reported by \citet{milisavljevic12} for their spectrum taken in October 2010, just over four years earlier.  This continues the gradual decline they noted from that of $13 \pm 2 \times 10^{-16} \FLUX$ measured by \citet{fesen99b} in November 1997, which in turn represented a fading of $\sim 25\%$ from the levels observed in the early 1990s. A similar fading of broad lines with time has been observed in M83 for SN 1957D \citep{long12}.

SN\,2004et, classified as Type IIP, was also well observed early on and as it made the transition to its nebular stage \citep[e.g.,][]{sahu06, maguire10, jerkstrand12}. The latest-time published spectra are by \cite{kotak09}, which extend the observations to just over 3 years post-explosion.  The spectrum in Fig.~\ref{historical_SNe} shows its recovery at an age of just over 10 years.  The most prominent feature is a very broad, asymmetric blend of \ha, \oi, and possibly \sii\ and/or \nii.   \cite{kotak09} observed a similar ``box-like" feature in spectra taken at 2.6 and 3.1 yr post-explosion, and  they measured a full width at zero intensity of $\sim 17,000 \kms$ in the spectrum at 3.1 yr.  The overall width in our 10.2 yr spectrum is similar or slightly broader, though it is not clear what physical significance to attach to this, since the feature results from blended lines.  Despite the blending, both the \ha\ and \oi\ contributions appear stronger on the blue side than the red, as in the case of  SN\,1980K\@.  \citet{kotak09} also note the presence of a narrow \ha\ component.  Such a component is also present in our spectrum, (Fig.~\ref{historical_SNe}); however, the two-dimensional spectrum from which the 1-D one was extracted shows this narrow \ha\ extending well beyond the broad components in the spatial direction, hence it is not clear that it is associated with SN\,2004et itself.

Also present in our spectrum is a fainter broad feature that is almost certainly \oiii, and a strong  feature at $\sim$7150 - 7400 \AA\ that is probably a blend of (primarily)  [\ion{Fe}{2}] $\lambda 7155$ and [\ion{Ca}{2}] $\lambda\lambda 7291, 7324$, features that were prominent in its late nebular spectra \citep{sahu06, maguire10}.  \oii\,$\lambda 7325$ may also be included in this blend.  (Unfortunately, this feature extends beyond the red end of our spectrum, making it harder to identify, but it is also present in the late Kotak et al.\ spectra, with a profile similar to the \ha/\oi\ feature.)  The \ha\ line flux is difficult to measure because it is so broad as to to be blended with \oi\ and, possibly, \sii; furthermore, the continuum level is also uncertain.  Our best flux estimate for \ha\ is $7 \pm 2 \times 10^{-16} \FLUX$.  Estimating the flux at 3.1 yr from Fig.\ 4 of \citet{kotak09}, the \ha\ flux at age 3.1 yr was $\sim 1\times 10^{-15} \FLUX$, so it appears to have faded very slightly over the intervening seven years.

The broad, asymmetric line profiles of both these SNe, stronger on the blue side than the red, are typical of the optical emission from other decades-old core-collapse SNe.  \citet{milisavljevic12} show several examples, and attribute the emission to the interaction between fast SN ejecta and the circumstellar shell from the progenitor star, as did \citet{kotak09} for SN\,2004et.  The predominence of blue-shifted over red-shifted emission may well result from the early formation of dust in cooling ejecta, resulting in greater absorption of emission from the far side of the expanding shell as it tries to make its way through the newly formed dust \citep[][and references therein]{milisavljevic12}.

The spectra of both SN\,1980K and SN\,2004et are also quite  similar to the object B12-174a identified in our similar survey of M83 \citep{blair15}. The main difference is that for B12-174a the SN was not observed, even though its inferred age is $<$100 years.  All these objects form a transitional class between ``old SNe" and mature SNRs.  SN1957D in M83 also shows broad lines, but only for oxygen, and the line intensities have dropped significantly over $\sim$40 years \citep[][and references therein]{long12}.  These differences may be due to differing progenitor types, differing local ISM conditions, or both.  \citep[The \oiii\ lines would appear relatively stronger in both  SN\,1980K and SN\,2004et if these were dereddened, with $E(B-V) = 0.41$,][but even so they would not be nearly so O-dominated as SN\,1957D.]{fesen99b, sahu06} Since there are so few objects in this transitional class, these objects bear watching for temporal changes that should happen on observable time scales.  Such observations could illuminate this poorly understood phase of SNR evolution.

The fact that none of the other seven historical SNe in \gal, ranging in age from 3 to 94 years, were detected is noteworthy.  All those with well-determined SN classifications are ones that result from core-collapse SNe, and thus should have produced several $M_{\sun}$ of high-velocity ejecta---the scenario responsible for ejecta-dominated SNRs like Cas A, or SN\,1957D in M83.  Furthermore, since \gal\ is such a champion producer of SNe, it is reasonable to expect the remnants from dozens of core-collapse SNe younger than 1000 yr to be located there.  The fact that so few are detected as broad-line, ejecta-dominated remnants is similar to the case of M83, host to six  SNe in the past century (or seven if we include B12-174a) and hence also expected to have far more young SNRs than are detected.  \citet{winkler17} concluded that many of the SNRs are likely expanding into high-density environments, where remnants evolve rapidly to the point that they are dominated by swept-up material rather than by ejecta. At the other extreme, other SNe may have exploded in regions where earlier SNRs have evacuated the surrounding region, resulting in extremely faint SNRs.  It would seem that the situation is similar in \gal.

\section{Summary \label{sec:summary}}

We have carried out a new optical search for SNRs in \gal\ using interference filters to identify emission nebulae that have elevated \sii:\ha\ ratios compared to \hii\ regions.  We recovered all of the SNRs that had been identified by MF97. Of the 147 SNR candidates we identified, we obtained spectra of 102, and spectroscopically confirmed  89 these based on elevated \sii:\ha\ ratios.  There are 45 candidates  without spectra and 17 candidates with spectra that show spectroscopic \sii:\ha\ ratios less than the canonical value of 0.4 for regarding an emission nebula as a confirmed optical SNR; many of these are in regions of \hii\ contamination that complicates spectral extraction.  Given the uncertainties, we have chosen to retain all 147  objects as SNR candidates, though only those 89 with high ratios can be regarded as confirmed.  

Very few of the candidates are detected as SNRs at other wavelengths.  Only seven are among the 35 radio SNRs identified by \cite{lacey01}, most likely due to the limited sensitivity of the radio survey. Similarly only eight candidates have X-ray counterparts, which we attribute to a combination of higher absorption along the line of sight to \gal\ compared to some other galaxies at comparable distances, e.g. M83, and to the 
lower exposure times for the {\em Chandra} study of \gal\ than for these other galaxies. 

We also inspected our images for evidence of emission at the sites of historical SNe in \gal\ and obtained spectra of the only two for which emission was apparent: SN\,1980K and SN\,2004et.  Both show the broad, asymmetric lines that are typical of very young SNRs, possibly caused by the interaction between fast SN ejecta and  circumstellar shells from the progenitors to these core-collapse SNe. Newly formed dust in cooling ejecta could then absorb light from the far side to produce the asymmetric profiles.  Although SN1980K is well-known as one of an unusual group of SNe that continue to be observable long after its explosion, the most recent (published) spectrum  of SN2004et was taken  3.1 years after its outburst \citep{kotak09}.  Our spectrum indicates that this object is still strongly interacting with circumstellar material to produce optical emission 10+ years after the explosion.

Much more work is needed to fully characterize the SNR population of \gal, some of which we are currently working on.  These include {\em HST} studies in the optical to measure diameters and identify additional small diameter objects in crowded regions, infrared \feii\ 1.64 $\mu$m studies to identify SNRs in dusty regions or buried in complex \ha\ emission, and deeper radio studies to find and characterize the radio counterparts of the optical SNR population. 

\acknowledgments

Our WIYN images were obtained at Kitt Peak National Observatory of the National Optical Astronomy Observatories (NOAO Prop. ID 11A-0110; PI: Winkler), which is operated by the Association of Universities for Research in Astronomy (AURA) under a cooperative agreement with the National Science Foundation. The spectra were obtained at the Gemini Observatory (Gemini Prop. IDs GN-2014A-Q-84,GN-2014B-Q-83, GN-2015B-Q-91; PI: Winkler), which is operated by the Association of Universities for Research in Astronomy, Inc., under a cooperative agreement with the NSF on behalf of the Gemini partnership: the National Science Foundation (United States), National Research Council (Canada), CONICYT (Chile), Ministerio de Ciencia, Tecnolog\'{i}a e Innovaci\'{o}n Productiva (Argentina), Minist\'{e}rio da Ci\^{e}ncia, Tecnologia e Inova\c{c}\~{a}o (Brazil), and Korea Astronomy and Space Science Institute (Republic of Korea).  Partial support for the analysis of the data was provided by NASA through grant number HST-GO-14638 from the Space Telescope Science Institute, which is operated by AURA, Inc., under NASA contract NAS 5-26555. PFW acknowledges additional support from the NSF through grant AST-1714281.  WPB acknowledges partial support from the JHU Center for Astrophysical Sciences.  We are grateful to the anonymous referee for making suggestions that have, we hope, improved this paper.

\vspace{5mm}
\facilities{NOAO:WIYN, Gemini:GMOS}


\software{astropy \citep{astropy} 
          }

\pagebreak

\bibliographystyle{aasjournal}

\bibliography{snr}

\clearpage


\begin{deluxetable}{ccrr}
\tablewidth{0pt}
\tablecaption{WIYN Imaging Observations of NGC\,6946}

\tablehead{
\colhead {} & \multicolumn{2}{c}{Filter} & \colhead {}\\ 
\cline{2-3}  
\colhead{Designation} &
\colhead{$\rm \lambda_{c}$(\AA)} &
\colhead{$\Delta \lambda$(\AA)\tablenotemark{a}} &
\colhead {Exposure (s)} 
}

\startdata
\oiii &  5010 &60\phn\phn & $11\times800$ \phn\phn  \\
Green Continuum  & 5127& 100\phn\phn&$11\times500$ \phn\phn \\
H$\alpha$  & 6563 & 27\phn\phn    & $10\times800$ \phn\phn \\
\sii\tablenotemark{b} & 6723 & 63\phn\phn & $10\times800$ \phn\phn  \\
Red Continuum  & 6840& 93\phn\phn &$10\times600$ \phn\phn \\
\enddata

\tablenotetext{a}{Full width at half maximum in the WIYN f/6.3 beam.}
\tablenotetext{b}{WIYN Observatory filter W037; other filters are PFW custom.}

\label{imaging_obsns}
\end{deluxetable}

\startlongtable

\clearpage

\begin{deluxetable}{rccrrrccl}
\rotate\tablecaption{SNR Candidates in NGC6946 }
\tablehead{\colhead{Source} & 
 \colhead{RA} & 
 \colhead{Dec} & 
 \colhead{GDC} & 
 \colhead{H$\alpha$~Flux$^a$} & 
 \colhead{[SII]:H$\alpha$} & 
 \colhead{Spectrum} & 
 \colhead{Confirmed} & 
 \colhead{Other~Names} 
\\
\colhead{~} & 
 \colhead{(2000)} & 
 \colhead{(2000)} & 
 \colhead{(kpc)} & 
 \colhead{~} & 
 \colhead{~} & 
 \colhead{~} & 
 \colhead{~} & 
 \colhead{~} 
}
\tabletypesize{\scriptsize}
\tablewidth{0pt}\startdata
L19-001 &  20:34:15.00 &  60:10:44.3  &  10.4 &  52 &  0.21 &  05.18 &  no &  -- \\ 
L19-002 &  20:34:15.48 &  60:07:31.6  &  9.6 &  64 &  0.34 &  -- &  -- &  -- \\ 
L19-003 &  20:34:15.78 &  60:08:26.0  &  9.2 &  216 &  1.14 &  -- &  -- &  -- \\ 
L19-004 &  20:34:16.41 &  60:08:27.3  &  9.0 &  33 &  0.61 &  02.25 &  no &  -- \\ 
L19-005 &  20:34:16.68 &  60:07:30.8  &  9.3 &  120 &  0.42 &  08.17 &  no &  -- \\ 
L19-006 &  20:34:17.54 &  60:10:58.3  &  10.1 &  97 &  0.66 &  05.09 &  yes &  -- \\ 
L19-007 &  20:34:17.95 &  60:10:00.4  &  9.1 &  92 &  0.49 &  02.10 &  yes &  -- \\ 
L19-008 &  20:34:18.39 &  60:10:47.3  &  9.7 &  540 &  0.33 &  -- &  -- &  -- \\ 
L19-009 &  20:34:18.84 &  60:11:08.9  &  10.0 &  33 &  0.86 &  05.01 &  yes &  -- \\ 
L19-010 &  20:34:19.17 &  60:08:57.5  &  8.3 &  251 &  0.40 &  02.21 &  yes &  -- \\ 
L19-011 &  20:34:20.60 &  60:09:06.8  &  8.0 &  56 &  0.52 &  02.11 &  yes &  -- \\ 
L19-012 &  20:34:21.96 &  60:08:57.8  &  7.6 &  90 &  0.50 &  -- &  -- &  -- \\ 
L19-013 &  20:34:22.70 &  60:06:13.4  &  9.4 &  15 &  0.82 &  08.01 &  yes &  -- \\ 
L19-014 &  20:34:23.38 &  60:08:18.7  &  7.3 &  96 &  0.62 &  02.01 &  yes &  MF-01; \\ 
L19-015 &  20:34:23.39 &  60:11:35.3  &  9.6 &  17 &  0.87 &  05.19 &  yes &  -- \\ 
L19-016 &  20:34:24.43 &  60:11:25.8  &  9.1 &  169 &  0.41 &  05.10 &  yes &  -- \\ 
L19-017 &  20:34:24.93 &  60:09:46.5  &  7.2 &  286 &  0.31 &  02.22 &  no &  -- \\ 
L19-018 &  20:34:25.37 &  60:08:56.4  &  6.7 &  65 &  0.39 &  -- &  -- &  -- \\ 
L19-019 &  20:34:26.00 &  60:11:10.5  &  8.4 &  103 &  0.65 &  05.02 &  yes &  MF-02; \\ 
L19-020 &  20:34:26.06 &  60:13:22.8  &  12.2 &  17 &  0.60 &  -- &  -- &  -- \\ 
L19-021 &  20:34:26.17 &  60:10:11.9  &  7.2 &  94 &  0.41 &  -- &  -- &  -- \\ 
L19-022 &  20:34:27.65 &  60:11:12.2  &  8.1 &  45 &  0.60 &  -- &  -- &  -- \\ 
L19-023 &  20:34:28.22 &  60:11:37.9  &  8.7 &  5 &  1.77 &  -- &  -- &  -- \\ 
L19-024 &  20:34:28.32 &  60:13:21.9  &  11.8 &  53 &  0.77 &  -- &  -- &  -- \\ 
L19-025 &  20:34:28.33 &  60:07:04.2  &  7.2 &  20 &  0.96 &  08.02 &  yes &  -- \\ 
L19-026 &  20:34:28.40 &  60:08:09.5  &  6.2 &  35 &  0.61 &  -- &  -- &  -- \\ 
L19-027 &  20:34:28.44 &  60:07:33.4  &  6.7 &  17 &  0.63 &  -- &  -- &  -- \\ 
L19-028 &  20:34:28.86 &  60:07:45.4  &  6.4 &  215 &  0.34 &  02.18 &  no &  -- \\ 
L19-029 &  20:34:29.17 &  60:10:51.1  &  7.3 &  12 &  1.26 &  -- &  -- &  F08-08; \\ 
L19-030 &  20:34:30.13 &  60:10:24.4  &  6.5 &  9 &  0.71 &  05.20 &  yes &  -- \\ 
L19-031 &  20:34:31.67 &  60:10:28.0  &  6.2 &  78 &  0.65 &  05.05 &  yes &  -- \\ 
L19-032 &  20:34:32.60 &  60:10:27.9  &  6.0 &  81 &  0.47 &  05.05 &  no &  -- \\ 
L19-033 &  20:34:33.05 &  60:11:25.7  &  7.4 &  134 &  0.49 &  05.11 &  yes &  -- \\ 
L19-034 &  20:34:33.31 &  60:09:46.7  &  5.1 &  13 &  1.12 &  -- &  -- &  -- \\ 
L19-035 &  20:34:33.65 &  60:09:52.0  &  5.1 &  14 &  1.32 &  -- &  -- &  MF-03; \\ 
L19-036 &  20:34:33.85 &  60:09:25.0  &  4.7 &  81 &  0.97 &  02.02 &  yes &  MF-04; \\ 
L19-037 &  20:34:36.63 &  60:11:34.4  &  7.0 &  186 &  0.44 &  05.03 &  yes &  -- \\ 
L19-038 &  20:34:37.38 &  60:07:15.0  &  5.4 &  42 &  0.66 &  02.03 &  yes &  -- \\ 
L19-039 &  20:34:37.44 &  60:11:31.4  &  6.8 &  36 &  0.77 &  04.01 &  yes &  -- \\ 
L19-040 &  20:34:37.76 &  60:08:52.6  &  3.6 &  24 &  0.90 &  08.07 &  yes &  MF-05; \\ 
L19-041 &  20:34:37.81 &  60:11:54.4  &  7.4 &  37 &  0.91 &  05.04 &  yes &  MF-06; \\ 
L19-042 &  20:34:37.98 &  60:07:22.3  &  5.1 &  18 &  1.31 &  02.04 &  yes &  MF-07; \\ 
L19-043 &  20:34:38.36 &  60:06:09.4  &  7.3 &  130 &  0.47 &  -- &  -- &  -- \\ 
L19-044 &  20:34:38.90 &  60:06:57.7  &  5.7 &  81 &  0.53 &  08.08 &  yes &  -- \\ 
L19-045 &  20:34:39.15 &  60:09:19.0  &  3.3 &  405 &  0.32 &  -- &  -- &  -- \\ 
L19-046 &  20:34:39.19 &  60:08:13.9  &  3.7 &  44 &  0.58 &  02.05 &  yes &  -- \\ 
L19-047 &  20:34:39.65 &  60:07:26.0  &  4.8 &  2 &  2.50 &  -- &  -- &  -- \\ 
L19-048 &  20:34:40.63 &  60:06:53.5  &  5.7 &  80 &  0.40 &  08.09 &  yes &  -- \\ 
L19-049 &  20:34:40.73 &  60:08:34.0  &  3.1 &  46 &  0.53 &  02.23 &  yes &  -- \\ 
L19-050 &  20:34:41.02 &  60:05:57.9  &  7.5 &  12 &  1.02 &  -- &  -- &  -- \\ 
L19-051 &  20:34:41.32 &  60:11:13.0  &  5.5 &  23 &  0.73 &  04.21 &  yes &  -- \\ 
L19-052 &  20:34:41.32 &  60:04:54.9  &  9.7 &  82 &  0.43 &  -- &  -- &  -- \\ 
L19-053 &  20:34:41.53 &  60:11:30.0  &  6.1 &  67 &  0.47 &  05.21 &  yes &  -- \\ 
L19-054 &  20:34:41.93 &  60:05:50.0  &  7.8 &  103 &  0.44 &  08.03 &  yes &  -- \\ 
L19-055 &  20:34:42.44 &  60:09:16.0  &  2.5 &  6 &  1.87 &  02.13 &  yes &  -- \\ 
L19-056 &  20:34:43.08 &  60:11:39.4  &  6.2 &  82 &  0.40 &  04.11 &  no &  -- \\ 
L19-057 &  20:34:43.32 &  60:10:11.1  &  3.3 &  187 &  0.44 &  -- &  -- &  -- \\ 
L19-058 &  20:34:43.53 &  60:07:51.7  &  3.5 &  25 &  0.67 &  -- &  -- &  -- \\ 
L19-059 &  20:34:43.97 &  60:08:24.4  &  2.6 &  54 &  0.62 &  02.14 &  yes &  MF-08; \\ 
L19-060 &  20:34:44.61 &  60:08:17.3  &  2.7 &  63 &  0.37 &  02.15 &  yes &  -- \\ 
L19-061 &  20:34:45.13 &  60:12:36.4  &  8.0 &  9 &  1.31 &  04.12 &  yes &  -- \\ 
L19-062 &  20:34:45.67 &  60:07:21.2  &  4.3 &  196 &  0.35 &  02.24 &  yes &  -- \\ 
L19-063 &  20:34:46.92 &  60:12:19.4  &  7.2 &  35 &  0.68 &  04.22 &  yes &  -- \\ 
L19-064 &  20:34:47.19 &  60:08:20.2  &  2.2 &  79 &  0.48 &  08.10 &  yes &  -- \\ 
L19-065 &  20:34:47.37 &  60:08:22.7  &  2.1 &  109 &  0.63 &  02.09 &  yes &  -- \\ 
L19-066 &  20:34:47.75 &  60:09:58.7  &  2.1 &  57 &  0.79 &  04.13 &  yes &  L97-34; \\ 
L19-067 &  20:34:48.09 &  60:07:50.5  &  3.2 &  97 &  0.44 &  08.11 &  yes &  -- \\ 
L19-068 &  20:34:48.64 &  60:09:24.4  &  1.0 &  159 &  0.44 &  07.01 &  yes &  -- \\ 
L19-069 &  20:34:48.72 &  60:08:23.4  &  2.0 &  138 &  0.41 &  01.01 &  yes &  -- \\ 
L19-070 &  20:34:49.66 &  60:07:37.0  &  3.6 &  60 &  0.50 &  03.10 &  yes &  -- \\ 
L19-071 &  20:34:49.80 &  60:09:41.3  &  1.2 &  69 &  0.40 &  -- &  -- &  -- \\ 
L19-072 &  20:34:49.95 &  60:07:53.5  &  3.0 &  50 &  0.54 &  06.10 &  yes &  -- \\ 
L19-073 &  20:34:50.02 &  60:09:43.3  &  1.3 &  86 &  0.49 &  -- &  -- &  -- \\ 
L19-074 &  20:34:50.36 &  60:09:45.2  &  1.3 &  79 &  0.38 &  02.16 &  yes &  -- \\ 
L19-075 &  20:34:50.37 &  60:09:51.8  &  1.5 &  579 &  0.24 &  -- &  -- &  -- \\ 
L19-076 &  20:34:50.80 &  60:07:48.4  &  3.2 &  159 &  0.31 &  03.11 &  yes &  F08-43;B14-20; \\ 
L19-077 &  20:34:50.94 &  60:10:20.9  &  2.6 &  3662 &  0.29 &  -- &  -- &  L97-48;F08-45; \\ 
L19-078 &  20:34:51.29 &  60:05:20.4  &  8.7 &  227 &  0.44 &  -- &  -- &  -- \\ 
L19-079 &  20:34:51.45 &  60:07:39.3  &  3.5 &  116 &  0.62 &  07.11 &  yes &  MF-09;L97-51; \\ 
L19-080 &  20:34:51.57 &  60:09:09.2  &  0.2 &  79 &  0.74 &  02.06 &  yes &  MF-10;F08-47; \\ 
L19-081 &  20:34:51.66 &  60:09:57.2  &  1.6 &  86 &  0.47 &  01.02 &  no &  -- \\ 
L19-082 &  20:34:52.47 &  60:07:28.2  &  4.0 &  39 &  0.89 &  03.12 &  yes &  MF-11; \\ 
L19-083 &  20:34:52.51 &  60:10:01.9  &  1.8 &  69 &  0.70 &  02.07 &  yes &  -- \\ 
L19-084 &  20:34:52.56 &  60:10:52.3  &  3.7 &  187 &  0.47 &  04.15 &  yes &  -- \\ 
L19-085 &  20:34:53.09 &  60:08:14.1  &  2.3 &  10 &  1.24 &  07.24 &  yes &  -- \\ 
L19-086 &  20:34:53.71 &  60:07:13.9  &  4.6 &  86 &  0.64 &  02.08 &  yes &  L97-68; \\ 
L19-087 &  20:34:54.31 &  60:11:03.4  &  4.0 &  33 &  0.96 &  04.02 &  yes &  MF-12; \\ 
L19-088 &  20:34:54.41 &  60:10:55.9  &  3.8 &  10 &  1.38 &  01.03 &  yes &  -- \\ 
L19-089 &  20:34:54.55 &  60:05:08.6  &  9.3 &  178 &  0.63 &  08.16 &  yes &  -- \\ 
L19-090 &  20:34:54.80 &  60:10:06.8  &  2.0 &  12 &  1.25 &  02.17 &  yes &  -- \\ 
L19-091 &  20:34:54.87 &  60:10:34.6  &  3.0 &  56 &  0.64 &  07.12 &  yes &  -- \\ 
L19-092 &  20:34:55.62 &  60:11:13.7  &  4.4 &  43 &  0.51 &  -- &  -- &  -- \\ 
L19-093 &  20:34:55.90 &  60:07:49.2  &  3.5 &  142 &  0.50 &  03.02 &  yes &  MF-13; \\ 
L19-094 &  20:34:56.58 &  60:08:19.9  &  2.5 &  88 &  0.76 &  01.04 &  yes &  F08-53; \\ 
L19-095 &  20:34:57.81 &  60:08:10.1  &  3.0 &  71 &  0.71 &  01.05 &  yes &  MF-14;B14-25; \\ 
L19-096 &  20:34:58.49 &  60:08:01.8  &  3.3 &  9 &  1.39 &  07.13 &  yes &  -- \\ 
L19-097 &  20:35:00.31 &  60:11:46.0  &  5.8 &  201 &  0.62 &  04.03 &  yes &  MF-15; \\ 
L19-098 &  20:35:00.72 &  60:11:30.9  &  5.3 &  1184 &  0.95 &  01.06 &  yes &  MF-16;L97-85;F08-63;B14-29; \\ 
L19-099 &  20:35:01.15 &  60:12:00.1  &  6.3 &  44 &  0.57 &  04.04 &  yes &  MF-17; \\ 
L19-100 &  20:35:02.24 &  60:11:05.2  &  4.6 &  274 &  0.48 &  01.07 &  yes &  -- \\ 
L19-101 &  20:35:02.38 &  60:06:31.5  &  7.0 &  188 &  0.57 &  03.04 &  yes &  MF-18; \\ 
L19-102 &  20:35:02.93 &  60:11:27.2  &  5.3 &  60 &  0.51 &  06.14 &  yes &  -- \\ 
L19-103 &  20:35:03.17 &  60:10:41.9  &  4.0 &  25 &  0.87 &  01.08 &  yes &  -- \\ 
L19-104 &  20:35:03.30 &  60:05:28.8  &  9.3 &  66 &  0.71 &  03.13 &  yes &  MF-19; \\ 
L19-105 &  20:35:03.59 &  60:06:23.4  &  7.4 &  79 &  0.41 &  -- &  -- &  -- \\ 
L19-106 &  20:35:04.06 &  60:11:15.6  &  5.1 &  14 &  1.33 &  04.16 &  yes &  -- \\ 
L19-107 &  20:35:04.19 &  60:11:18.5  &  5.2 &  48 &  0.72 &  -- &  -- &  -- \\ 
L19-108 &  20:35:04.22 &  60:09:53.5  &  3.2 &  43 &  0.64 &  06.15 &  no &  L97-88; \\ 
L19-109 &  20:35:04.27 &  60:06:52.1  &  6.5 &  11 &  1.13 &  03.14 &  yes &  -- \\ 
L19-110 &  20:35:05.00 &  60:05:32.9  &  9.3 &  48 &  0.50 &  -- &  -- &  -- \\ 
L19-111 &  20:35:05.63 &  60:10:00.8  &  3.6 &  47 &  0.60 &  06.16 &  yes &  MF-20; \\ 
L19-112 &  20:35:05.69 &  60:11:07.6  &  5.1 &  383 &  0.32 &  04.06 &  yes &  L97-95; \\ 
L19-113 &  20:35:06.89 &  60:07:58.4  &  5.0 &  40 &  0.59 &  03.15 &  yes &  -- \\ 
L19-114 &  20:35:06.96 &  60:09:57.0  &  3.9 &  98 &  0.51 &  04.07 &  yes &  -- \\ 
L19-115 &  20:35:07.07 &  60:05:57.3  &  8.8 &  247 &  0.36 &  -- &  -- &  -- \\ 
L19-116 &  20:35:08.80 &  60:06:03.0  &  8.8 &  63 &  0.45 &  -- &  -- &  MF-21; \\ 
L19-117 &  20:35:08.89 &  60:10:13.0  &  4.5 &  9 &  1.08 &  -- &  -- &  -- \\ 
L19-118 &  20:35:09.56 &  60:09:13.1  &  4.4 &  239 &  0.39 &  06.17 &  no &  -- \\ 
L19-119 &  20:35:09.61 &  60:12:30.0  &  8.0 &  124 &  0.73 &  -- &  -- &  MF-22; \\ 
L19-120 &  20:35:09.87 &  60:06:13.3  &  8.6 &  19 &  0.82 &  -- &  -- &  -- \\ 
L19-121 &  20:35:10.22 &  60:06:26.7  &  8.3 &  83 &  0.49 &  03.16 &  yes &  -- \\ 
L19-122 &  20:35:10.54 &  60:06:41.3  &  7.9 &  23 &  0.78 &  -- &  -- &  -- \\ 
L19-123 &  20:35:10.63 &  60:10:40.9  &  5.3 &  190 &  0.39 &  01.11 &  yes &  -- \\ 
L19-124 &  20:35:10.89 &  60:08:56.9  &  4.9 &  825 &  0.33 &  06.04 &  no &  F08-74; \\ 
L19-125 &  20:35:11.04 &  60:08:27.1  &  5.3 &  70 &  0.49 &  01.12 &  yes &  -- \\ 
L19-126 &  20:35:11.45 &  60:11:11.9  &  6.1 &  110 &  0.57 &  04.17 &  yes &  -- \\ 
L19-127 &  20:35:11.60 &  60:07:41.2  &  6.4 &  183 &  0.51 &  03.05 &  yes &  MF-23; \\ 
L19-128 &  20:35:11.90 &  60:09:28.6  &  5.0 &  20 &  0.88 &  06.18 &  yes &  -- \\ 
L19-129 &  20:35:11.94 &  60:04:03.7  &  13.3 &  342 &  0.34 &  -- &  -- &  -- \\ 
L19-130 &  20:35:12.25 &  60:06:37.6  &  8.3 &  69 &  0.57 &  -- &  -- &  -- \\ 
L19-131 &  20:35:12.62 &  60:09:09.7  &  5.2 &  60 &  0.62 &  01.13 &  yes &  -- \\ 
L19-132 &  20:35:13.62 &  60:08:58.9  &  5.5 &  111 &  0.54 &  07.25 &  yes &  -- \\ 
L19-133 &  20:35:14.44 &  60:07:12.7  &  7.7 &  9 &  1.10 &  07.18 &  yes &  -- \\ 
L19-134 &  20:35:16.52 &  60:07:50.1  &  7.3 &  11 &  0.77 &  -- &  -- &  -- \\ 
L19-135 &  20:35:16.93 &  60:11:05.4  &  7.0 &  56 &  0.86 &  01.15 &  yes &  MF-24; \\ 
L19-136 &  20:35:17.33 &  60:10:27.3  &  6.6 &  21 &  0.80 &  04.18 &  yes &  -- \\ 
L19-137 &  20:35:17.56 &  60:07:19.3  &  8.2 &  200 &  0.47 &  -- &  -- &  -- \\ 
L19-138 &  20:35:20.08 &  60:09:33.9  &  7.0 &  88 &  0.61 &  06.05 &  yes &  F08-82; \\ 
L19-139 &  20:35:20.80 &  60:09:52.7  &  7.2 &  16 &  1.17 &  -- &  -- &  -- \\ 
L19-140 &  20:35:21.11 &  60:08:44.1  &  7.6 &  203 &  0.63 &  06.06 &  yes &  MF-25; \\ 
L19-141 &  20:35:23.02 &  60:08:21.2  &  8.3 &  200 &  0.38 &  01.18 &  yes &  -- \\ 
L19-142 &  20:35:23.66 &  60:08:47.7  &  8.2 &  129 &  0.44 &  07.20 &  no &  -- \\ 
L19-143 &  20:35:24.22 &  60:07:42.5  &  9.2 &  124 &  0.41 &  03.17 &  no &  -- \\ 
L19-144 &  20:35:24.66 &  60:06:57.2  &  10.3 &  18 &  0.80 &  -- &  -- &  -- \\ 
L19-145 &  20:35:25.24 &  60:07:26.9  &  9.8 &  299 &  0.38 &  -- &  -- &  -- \\ 
L19-146 &  20:35:25.51 &  60:07:51.3  &  9.4 &  57 &  0.67 &  -- &  -- &  MF-26; \\ 
L19-147 &  20:35:26.11 &  60:08:43.0  &  8.8 &  202 &  0.65 &  01.19 &  yes &  MF-27; \\ 
\enddata 
\tablenotetext{a}{ H$\alpha$ Flux is in units of 10$^{-17}$ ergs cm$^{-2}$ s$^{-1}$.}
\tablenotetext{b}{MF = \cite{matonick97};  B14 = \cite{bruursema14}; F08 = \cite{fridriksson08}; L97 = \cite{lacey97}}
\label{table_candidates}
\end{deluxetable}

\pagebreak

\begin{deluxetable}{ccc}
\tablewidth{0pt}
\tablecaption{Gemini-N/GMOS Multi-Object Spectroscopy Observations of NGC\,6946}

\tablehead{
\colhead{Mask No.} &
\colhead{Date (UT)}  &
\colhead{Total Exposure (s)\tablenotemark{a} }
}

\startdata
 1 & 30 Jul 2014 & 3 CWLs $\times 2 \times 1800$  \\[2pt]
 2 & 24 Sep 2014 & 3 CWLs $\times 2 \times 1800$ \\[2pt]
3 & 25-30 Sep 2014 & 3 CWLs $\times 2 \times 1800$  \\[2pt]
4 & 26 Oct, 19 Nov 2014 & 3 CWLs $\times 2 \times 1800$  \\[2pt]
5 & 21-27 Nov 2014 & 3 CWLs $\times 2 \times 1800$  \\[2pt]
6\tablenotemark{b}  & 14-17 Dec 2014 & 2 CWLs $\times 2 \times 1800$  \\[2pt]
7 & 14 Sep 2015 & 3 CWLs $\times 3 \times 1200$  \\[2pt]
8 & 20 Sep\,-\,19 Oct 2015 & 4 CWLs $\times 3 \times 1000$  \\[2pt]
\enddata

\tablenotetext{a}{Number of different Central Wavelength Settings $\times$ number of exposures at each CWL $\times$ individual exposure time.}
\tablenotetext{b}{Observations for mask 6, done late in the 2014B semester, were incomplete.  Many of the same objects were re-observed with mask 7 in 2015B.}

\label{obs_log}
\end{deluxetable}

\clearpage

\startlongtable

\begin{deluxetable}{rrrrrrrrrrrr}
\rotate\tablecaption{Emission line fluxes of SNR candidates$^{a,b,c}$ }
\tablehead{\colhead{Source} & 
 \colhead{H$\alpha$~flux} & 
 \colhead{H$\beta$} & 
 \colhead{[OIII]5007} & 
 \colhead{[OI]6300} & 
 \colhead{H$\alpha$} & 
 \colhead{[NII]6583} & 
 \colhead{[SII]6716} & 
 \colhead{[SII]6731} & 
 \colhead{[SII]:H$\alpha$} & 
 \colhead{[SII]6716:6731} & 
 \colhead{FWHM} 
}
\tabletypesize{\scriptsize}
\tablewidth{0pt}\startdata
L19-001 &  183 &  76 &  -- &  -- &  300 &  38 &  33 &  30 &  0.21 &  1.10 &  6.8 \\ 
L19-004 &  68 &  53 &  -- &  -- &  300 &  $\sim$93 &  69 &  47 &  0.39 &  1.47 &  7.1 \\ 
L19-005 &  216 &  80 &  $\sim$182 &  -- &  300 &  61 &  65 &  43 &  0.36 &  1.51 &  8.3 \\ 
L19-006 &  181 &  79 &  -- &  -- &  300 &  102 &  170 &  113 &  0.94 &  1.50 &  8.4 \\ 
L19-007 &  123 &  54 &  87 &  -- &  300 &  149 &  84 &  58 &  0.47 &  1.45 &  7.4 \\ 
L19-009 &  122 &  29 &  81 &  40 &  300 &  114 &  180 &  138 &  1.06 &  1.30 &  7.6 \\ 
L19-010 &  57 &  $\sim$49 &  -- &  $\sim$38 &  300 &  $\sim$111 &  79 &  53 &  0.44 &  1.49 &  9.4 \\ 
L19-011 &  28 &  $\sim$42 &  $\sim$202 &  $\sim$18 &  300 &  $\sim$134 &  $\sim$102 &  $\sim$78 &  $\sim$0.60 &   $\sim$1.31 &  8.5 \\ 
L19-013 &  19 &  $\sim$61 &  $\sim$165 &  $\sim$76 &  300 &  $\sim$148 &  $\sim$132 &  $\sim$127 &  $\sim$0.86 &   $\sim$1.04 &  7.0 \\ 
L19-014 &  132 &  38 &  108 &  40 &  300 &  164 &  123 &  89 &  0.71 &  1.38 &  7.4 \\ 
L19-015 &  52 &  -- &  146 &  -- &  300 &  139 &  117 &  112 &  0.76 &  1.04 &  9.6 \\ 
L19-016 &  102 &  -- &  32 &  36 &  300 &  135 &  149 &  103 &  0.84 &  1.45 &  6.1 \\ 
L19-017 &  248 &  31 &  15 &  73 &  300 &  95 &  49 &  29 &  0.26 &  1.69 &  7.6 \\ 
L19-019 &  85 &  36 &  139 &  209 &  300 &  180 &  239 &  171 &  1.37 &  1.40 &  6.5 \\ 
L19-025 &  21 &  -- &  $\sim$166 &  $\sim$81 &  300 &  $\sim$225 &  182 &  133 &  1.05 &  1.37 &  7.8 \\ 
L19-028 &  150 &  29 &  9 &  -- &  300 &  99 &  57 &  43 &  0.33 &  1.33 &  5.5 \\ 
L19-030 &  31 &  -- &  -- &  874 &  300 &  $\sim$168 &  177 &  121 &  0.99 &  1.46 &  6.2 \\ 
L19-031 &  179 &  69 &  110 &  67 &  300 &  217 &  189 &  134 &  1.08 &  1.41 &  7.6 \\ 
L19-032 &  501 &  55 &  $\sim$13 &  $\sim$6 &  300 &  119 &  64 &  49 &  0.38 &  1.31 &  7.4 \\ 
L19-033 &  157 &  75 &  -- &  -- &  300 &  179 &  132 &  85 &  0.72 &  1.55 &  7.2 \\ 
L19-036 &  89 &  35 &  56 &  81 &  300 &  296 &  218 &  161 &  1.26 &  1.35 &  7.7 \\ 
L19-037 &  43 &  -- &  -- &  $\sim$55 &  300 &  $\sim$142 &  243 &  167 &  1.37 &  1.46 &  4.0 \\ 
L19-038 &  28 &  $\sim$56 &  $\sim$67 &  $\sim$63 &  300 &  $\sim$224 &  173 &  115 &  0.96 &  1.50 &  6.1 \\ 
L19-039 &  78 &  $\sim$31 &  $\sim$76 &  $\sim$-357 &  300 &  240 &  189 &  140 &  1.09 &  1.35 &  7.1 \\ 
L19-040 &  20 &  -- &  $\sim$23 &  $\sim$60 &  300 &  $\sim$314 &  231 &  166 &  1.32 &  1.39 &  7.0 \\ 
L19-041 &  55 &  117 &  $\sim$177 &  -- &  300 &  $\sim$202 &  156 &  146 &  1.01 &  1.07 &  5.9 \\ 
L19-042 &  26 &  $\sim$39 &  $\sim$158 &  $\sim$89 &  300 &  390 &  240 &  175 &  1.38 &  1.37 &  7.3 \\ 
L19-044 &  15 &  $\sim$88 &  $\sim$106 &  $\sim$49 &  300 &  $\sim$180 &  $\sim$117 &  $\sim$69 &  $\sim$0.62 &   $\sim$1.70 &  6.7 \\ 
L19-046 &  31 &  $\sim$32 &  $\sim$92 &  $\sim$93 &  300 &  366 &  177 &  149 &  1.09 &  1.19 &  7.7 \\ 
L19-048 &  69 &  50 &  56 &  47 &  300 &  146 &  79 &  55 &  0.45 &  1.44 &  6.9 \\ 
L19-049 &  56 &  $\sim$35 &  $\sim$76 &  $\sim$13 &  300 &  185 &  95 &  65 &  0.53 &  1.46 &  7.5 \\ 
L19-051 &  45 &  -- &  182 &  242 &  300 &  303 &  191 &  147 &  1.13 &  1.30 &  7.5 \\ 
L19-053 &  239 &  64 &  -- &  20 &  300 &  103 &  84 &  63 &  0.49 &  1.33 &  6.5 \\ 
L19-054 &  88 &  56 &  -- &  -- &  300 &  107 &  79 &  56 &  0.45 &  1.41 &  6.7 \\ 
L19-055 &  9 &  $\sim$86 &  -- &  $\sim$65 &  300 &  $\sim$420 &  $\sim$227 &  $\sim$160 &  $\sim$1.29 &   $\sim$1.42 &  6.6 \\ 
L19-056 &  97 &  44 &  $\sim$21 &  -- &  300 &  112 &  62 &  44 &  0.35 &  1.41 &  7.9 \\ 
L19-059 &  62 &  $\sim$40 &  -- &  -- &  300 &  264 &  126 &  90 &  0.72 &  1.40 &  7.2 \\ 
L19-060 &  50 &  $\sim$42 &  $\sim$46 &  $\sim$49 &  300 &  205 &  82 &  66 &  0.49 &  1.24 &  7.2 \\ 
L19-061 &  60 &  -- &  -- &  -- &  300 &  177 &  198 &  146 &  1.15 &  1.36 &  7.1 \\ 
L19-062 &  85 &  47 &  $\sim$51 &  $\sim$10 &  300 &  131 &  83 &  60 &  0.48 &  1.38 &  5.7 \\ 
L19-063 &  116 &  52 &  $\sim$26 &  -- &  300 &  118 &  85 &  69 &  0.51 &  1.23 &  8.4 \\ 
L19-064 &  56 &  $\sim$26 &  $\sim$39 &  -- &  300 &  $\sim$173 &  105 &  70 &  0.58 &  1.50 &  6.5 \\ 
L19-065 &  48 &  $\sim$8 &  $\sim$19 &  $\sim$60 &  300 &  232 &  170 &  137 &  1.02 &  1.24 &  6.1 \\ 
L19-066 &  36 &  -- &  -- &  -- &  300 &  350 &  314 &  198 &  1.71 &  1.59 &  5.9 \\ 
L19-067 &  69 &  55 &  -- &  -- &  300 &  157 &  118 &  93 &  0.70 &  1.27 &  7.0 \\ 
L19-068 &  92 &  $\sim$30 &  47 &  39 &  300 &  270 &  140 &  105 &  0.82 &  1.33 &  6.7 \\ 
L19-069 &  222 &  35 &  42 &  26 &  300 &  188 &  92 &  72 &  0.54 &  1.28 &  8.1 \\ 
L19-070 &  49 &  62 &  127 &  -- &  300 &  270 &  118 &  86 &  0.68 &  1.37 &  7.9 \\ 
L19-072 &  27 &  -- &  $\sim$77 &  $\sim$40 &  300 &  323 &  203 &  113 &  1.06 &  1.80 &  6.4 \\ 
L19-074 &  24 &  $\sim$32 &  -- &  258 &  300 &  739 &  368 &  302 &  2.24 &  1.22 &  8.1 \\ 
L19-076 &  47 &  -- &  78 &  104 &  300 &  607 &  132 &  182 &  1.05 &  0.73 &  9.2 \\ 
L19-079 &  22 &  $\sim$18 &  $\sim$62 &  $\sim$54 &  300 &  $\sim$343 &  $\sim$135 &  $\sim$97 &  $\sim$0.77 &   $\sim$1.39 &  6.9 \\ 
L19-080 &  73 &  $\sim$13 &  158 &  78 &  300 &  637 &  185 &  178 &  1.21 &  1.04 &  9.5 \\ 
L19-081 &  99 &  $\sim$13 &  $\sim$47 &  $\sim$20 &  300 &  142 &  63 &  45 &  0.36 &  1.40 &  8.1 \\ 
L19-082 &  23 &  -- &  -- &  $\sim$111 &  300 &  $\sim$389 &  261 &  181 &  1.47 &  1.44 &  6.6 \\ 
L19-083 &  24 &  $\sim$32 &  $\sim$176 &  $\sim$87 &  300 &  629 &  234 &  182 &  1.39 &  1.29 &  7.7 \\ 
L19-084 &  119 &  51 &  24 &  41 &  300 &  140 &  109 &  76 &  0.62 &  1.43 &  7.8 \\ 
L19-085 &  37 &  -- &  $\sim$235 &  $\sim$43 &  300 &  493 &  182 &  41 &  0.74 &  4.44 &  8.6 \\ 
L19-086 &  28 &  -- &  $\sim$174 &  $\sim$58 &  300 &  393 &  245 &  128 &  1.24 &  1.91 &  7.5 \\ 
L19-087 &  100 &  42 &  46 &  121 &  300 &  263 &  204 &  156 &  1.20 &  1.31 &  7.1 \\ 
L19-088 &  24 &  $\sim$81 &  $\sim$271 &  $\sim$61 &  300 &  $\sim$232 &  216 &  157 &  1.24 &  1.38 &  5.4 \\ 
L19-089 &  11 &  -- &  -- &  -- &  300 &  $\sim$160 &  $\sim$138 &  $\sim$76 &  $\sim$0.71 &   $\sim$1.82 &  11.0 \\ 
L19-090 &  27 &  -- &  -- &  -- &  300 &  499 &  180 &  155 &  1.12 &  1.16 &  7.6 \\ 
L19-091 &  9 &  $\sim$84 &  -- &  -- &  300 &  $\sim$387 &  $\sim$332 &  $\sim$253 &  $\sim$1.95 &   $\sim$1.31 &  6.4 \\ 
L19-093 &  159 &  49 &  56 &  -- &  300 &  175 &  100 &  71 &  0.57 &  1.41 &  6.9 \\ 
L19-094 &  179 &  37 &  153 &  60 &  300 &  433 &  107 &  119 &  0.75 &  0.90 &  7.9 \\ 
L19-095 &  108 &  34 &  56 &  52 &  300 &  249 &  119 &  115 &  0.78 &  1.03 &  6.7 \\ 
L19-096 &  14 &  $\sim$95 &  $\sim$70 &  $\sim$145 &  300 &  $\sim$406 &  291 &  207 &  1.66 &  1.41 &  7.2 \\ 
L19-097 &  564 &  58 &  -- &  -- &  300 &  181 &  140 &  105 &  0.82 &  1.33 &  6.7 \\ 
L19-098 &  1351 &  -- &  518 &  97 &  300 &  276 &  160 &  153 &  1.04 &  1.05 &  7.8 \\ 
L19-099 &  171 &  53 &  $\sim$120 &  $\sim$15 &  300 &  149 &  116 &  91 &  0.69 &  1.27 &  6.7 \\ 
L19-100 &  61 &  51 &  -- &  -- &  300 &  $\sim$120 &  93 &  52 &  0.48 &  1.79 &  7.6 \\ 
L19-101 &  243 &  58 &  4 &  16 &  300 &  118 &  84 &  57 &  0.47 &  1.47 &  8.8 \\ 
L19-102 &  103 &  -- &  -- &  -- &  300 &  105 &  97 &  69 &  0.55 &  1.41 &  7.1 \\ 
L19-103 &  56 &  $\sim$34 &  86 &  60 &  300 &  298 &  149 &  110 &  0.86 &  1.35 &  6.9 \\ 
L19-104 &  120 &  53 &  134 &  -- &  300 &  96 &  104 &  73 &  0.59 &  1.42 &  8.8 \\ 
L19-106 &  73 &  -- &  129 &  55 &  300 &  217 &  193 &  141 &  1.11 &  1.37 &  6.8 \\ 
L19-108 &  330 &  23 &  -- &  -- &  300 &  127 &  58 &  43 &  0.34 &  1.35 &  6.0 \\ 
L19-109 &  18 &  $\sim$54 &  $\sim$282 &  -- &  300 &  $\sim$246 &  184 &  128 &  1.04 &  1.44 &  8.8 \\ 
L19-111 &  64 &  $\sim$17 &  -- &  50 &  300 &  190 &  120 &  99 &  0.73 &  1.21 &  8.7 \\ 
L19-112 &  258 &  45 &  98 &  35 &  300 &  170 &  116 &  90 &  0.69 &  1.29 &  6.8 \\ 
L19-113 &  31 &  -- &  -- &  $\sim$36 &  300 &  $\sim$158 &  104 &  82 &  0.62 &  1.27 &  6.1 \\ 
L19-114 &  311 &  29 &  9 &  14 &  300 &  150 &  95 &  72 &  0.56 &  1.32 &  6.7 \\ 
L19-118 &  863 &  37 &  -- &  8 &  300 &  116 &  52 &  37 &  0.29 &  1.41 &  7.3 \\ 
L19-121 &  92 &  52 &  $\sim$51 &  $\sim$24 &  300 &  $\sim$105 &  101 &  73 &  0.58 &  1.38 &  9.0 \\ 
L19-123 &  12 &  $\sim$31 &  $\sim$28 &  -- &  300 &  $\sim$239 &  $\sim$186 &  $\sim$133 &  $\sim$1.06 &   $\sim$1.40 &  6.0 \\ 
L19-124 &  2638 &  34 &  5 &  14 &  300 &  115 &  52 &  44 &  0.32 &  1.18 &  6.2 \\ 
L19-125 &  127 &  $\sim$23 &  24 &  39 &  300 &  135 &  97 &  73 &  0.57 &  1.33 &  6.2 \\ 
L19-126 &  80 &  $\sim$35 &  -- &  -- &  300 &  $\sim$103 &  89 &  60 &  0.50 &  1.48 &  7.2 \\ 
L19-127 &  242 &  44 &  89 &  35 &  300 &  147 &  97 &  80 &  0.59 &  1.21 &  8.0 \\ 
L19-128 &  40 &  -- &  65 &  133 &  300 &  323 &  -- &  -- &  -- &  1.00 &  6.9 \\ 
L19-131 &  60 &  $\sim$34 &  192 &  74 &  300 &  266 &  165 &  145 &  1.03 &  1.14 &  7.9 \\ 
L19-132 &  61 &  $\sim$49 &  -- &  -- &  300 &  155 &  105 &  82 &  0.62 &  1.28 &  7.6 \\ 
L19-133 &  7 &  -- &  $\sim$148 &  $\sim$122 &  300 &  $\sim$171 &  $\sim$215 &  $\sim$138 &  $\sim$1.18 &   $\sim$1.56 &  5.5 \\ 
L19-135 &  93 &  51 &  46 &  91 &  300 &  139 &  180 &  132 &  1.04 &  1.36 &  8.0 \\ 
L19-136 &  74 &  47 &  -- &  -- &  300 &  129 &  174 &  116 &  0.97 &  1.50 &  7.1 \\ 
L19-138 &  317 &  52 &  55 &  69 &  300 &  168 &  102 &  97 &  0.66 &  1.05 &  6.9 \\ 
L19-140 &  472 &  -- &  -- &  36 &  300 &  102 &  103 &  76 &  0.60 &  1.36 &  7.5 \\ 
L19-141 &  82 &  62 &  115 &  63 &  300 &  173 &  130 &  99 &  0.76 &  1.31 &  7.9 \\ 
L19-142 &  155 &  38 &  $\sim$116 &  $\sim$8 &  300 &  83 &  39 &  29 &  0.23 &  1.34 &  6.7 \\ 
L19-143 &  152 &  60 &  $\sim$23 &  -- &  300 &  89 &  58 &  41 &  0.33 &  1.41 &  8.1 \\ 
L19-147 &  271 &  55 &  11 &  16 &  300 &  112 &  106 &  74 &  0.60 &  1.43 &  8.4 \\ 
\enddata 
\tablenotetext{a}{ H$\alpha$ Flux is in units of 10$^{-17}$ ergs cm$^{-2}$ s$^{-1}$.}
\tablenotetext{b}{ Emission line strengths are listed relative to H$\alpha$ set to 300.}
\tablenotetext{c}{ FWHM is in \AA.}
\label{table_snr_spec}
\end{deluxetable}

\pagebreak

%

\end{document}